\documentclass{iopart}
\usepackage[latin1]{inputenc}
\usepackage{iopams}
\usepackage{color}
\usepackage{graphicx}

\newcommand{\ip}{\mathrm{I}_\mathrm{p}}
\newcommand{\up}{\mathrm{U}_\mathrm{p}}
\renewcommand\Re{\mathrm{Re}}
\renewcommand\Im{\mathrm{Im}}

\begin{document}

\title{Analytic quantum-interference conditions in Coulomb corrected photoelectron holography }
\author{A. S. Maxwell$^1$\thanks{andrew.maxwell.14@ucl.ac.uk},
	A. Al-Jawahiry$^1$, X. Y. Lai$^2$
	and C. Figueira de Morisson Faria$^1$\thanks{c.faria@ucl.ac.uk} }

\address{$^1$Department of Physics and Astronomy, University College London \\
	Gower Street London WC1E 6BT, United Kingdom\\$^2$State Key Laboratory of Magnetic Resonance and Atomic and Molecular Physics,
	Wuhan Institute of Physics and Mathematics, Chinese Academy of Sciences, Wuhan 430071, China }

\begin{abstract}
	We provide approximate analytic expressions for above-threshold ionization (ATI) transition probabilities and photoelectron angular distributions (PADs). These analytic expressions are more general than those existing in the literature and include the residual binding potential in the electron continuum propagation.
	They successfully reproduce the  ATI side lobes and specific holographic structures such as the near-threshold
	fan-shaped pattern and the spider-like structure that extends up to
	relatively high photoelectron energies.  We compare such expressions with the Coulomb quantum orbit strong-field approximation (CQSFA) and the full solution of the time-dependent Schr\"odinger equation for different driving-field frequencies and intensities, and provide an in-depth analysis of the physical mechanisms behind specific holographic structures. Our results shed additional light on what aspects of the CQSFA must be prioritized in order to obtain the key holographic features, and highlight the importance of forward scattered trajectories. Furthermore, we find that the holographic patterns change considerably for different field
	parameters, even if the Keldysh parameter is kept roughly the same. 
\end{abstract}


\maketitle

\section{Introduction}

\label{sec:intro}
Quantum interference in above-threshold ionization (ATI) photoelectron angular distributions (PADs) plays an important role in attosecond imaging and photoelectron holography. The physical picture that associates ionization events with electron trajectories \cite{Corkum1993} provides a very intuitive explanation, which can be used to disentangle how specific holographic patterns form. For a given final momentum, there are many pathways that the electron can follow, so that the corresponding transition amplitudes interfere. Photoelectron holography requires a probe and a reference signal, which are associated with two distinct types of orbits. These trajectories are either classical, or have a classical counterpart. For other types of trajectories, see, e.g., \cite{Wu2013a,Wu2013b,Zagoya2014,Symonds2015}.

There are several types of holographic structures, among them the fork- or spider-like pattern that forms near the polarization axis and extends up to high photoelectron energies, and a fishbone-like structure observed for molecular targets (for examples of such structures see \cite{Bian2011PRA}).
To the present
date, most expressions provided for the maxima and minima in holographic
structures either rely on drastic simplifications, or have a limited range
of validity. Classical models, in which the influence of the binding potential is considered upon rescattering, but neglected in the continuum propagation, have been widely used.   Coulomb-corrected conditions are only provided for specific
scattering angles, and consider mainly sub-barrier corrections \cite{Yan2012}. Still, such models have been able to reproduce key features. For instance, simplified models have shown that the spider-like structure stems from the interference between different types of deflected
trajectories leaving within the same half cycle \cite{Huismanset2010Science,Huismanset2012PRL,Hickstein2012,LiPRL2014}.

Nonetheless, the combined effect of the long-range binding potential and the external driving field on the electron trajectories is important. For instance, we have shown, using the Coulomb-quantum orbit strong-field approximation (CQSFA) \cite{Lai2015}, that the fan-shaped structure that forms near the ionization threshold results from the interference of trajectories that reach the detector directly with those that are forward-deflected by the core without undergoing hard collisions \cite{Lai2017}. The fan-like fringes are caused by the fact that the Coulomb potential distorts the deflected trajectories unequally for different scattering angles and electron momenta \cite{Lai2017,Maxwell2017}. Whilst this structure is widely studied and known to occur for  long-range potentials, in most cases methods have been used that hinder a direct statement about how these patterns \textit{form}. These include the numerical solution of the time-dependent Schr\"odinger equation for short- and long-range potentials, for which specific sets of orbits cannot be disentangled, or classical-trajectory methods, for which quantum interference does not occur \cite{Chen2006PRA,Arbo2006PRL,Arbo2008PRA}. Additionally, we have found that side lobes are already present in Coulomb corrected, single-orbit probability distributions \cite{Maxwell2017}, even if they are enhanced by constructive intra-cycle interference. This shows that the interference mechanism that leads to the spider is not the sole cause of the ATI side lobes.  In fact, they are also due to the Coulomb potential modifying the electron's tunneling probability. 

The above-stated examples show that Coulomb effects are under-estimated and poorly understood within photoelectron holography. However, oversimplifications do have the advantage of leading to intuitive analytic conditions that describe the key features in several holographic patterns.  This invites the following questions: Is it possible to  derive more general expressions than in previous models, which account for the Coulomb potential, but are transparent enough to highlight the key features? If so, what is their range of validity?

These questions will be addressed in the present work. In this publication, we will seek analytic expressions so that the fan- and spider-like holographic patterns are reproduced for a wide range of driving-field parameters. We will keep the same Keldysh parameter $%
\gamma =\sqrt{I_{p}/(2U_{p})}$, where $I_{p}$ and $U_{p}$ denote the
ionization potential and the ponderomotive energy, as, traditionally, it is a good indicator of the ionization dynamics. Furthermore, there is experimental evidence
that, for approximately the same $\gamma$, the spider-shaped structure becomes
more important with increasing wavelength, in detriment of the fan-shaped
pattern, and that the number of maxima in the fan changes \cite%
{Maharjan2006JPB}. As much as possible, we will justify the
behaviours encountered in terms of interfering trajectories. Throughout, we
will use the orbit classification introduced in \cite{Yan2010,Yan2012},
where a closely related Coulomb-corrected strong-field approximation (CCSFA)
is employed. This classification is based on the initial and final momentum
components parallel and perpendicular to the driving-field polarization, and
has been used in our previous publications \cite{Lai2015,Lai2017,Maxwell2017}.
It singles out four types of orbits, three of which were found to be relevant to the
parameter range in \cite{Lai2017,Maxwell2017}. Our results will be compared with the outcome of the CQSFA, and with the time-dependent Schr\"odinger equation (TDSE), which is solved using the freely available software Qprop \cite{Qprop}. We use atomic units throughout.

This work will complement and extend previous studies, in which we have provided a general expression encompassing inter- and intracycle interference \cite{Maxwell2017}, which could be factorized for monochromatic fields. We have also derived analytic expressions for inter-cycle interference valid in the presence of the Coulomb potential, and for single-orbit
transition amplitudes that justify the presence of side lobes. Intra-cycle interference, however, has mainly been discussed numerically and qualitatively  \cite{Lai2017,Maxwell2017}. We have also limited our investigation to near-infrared fields. 

This article is organized as follows. In Sec.~\ref{sec:theory} we state the full CQSFA expressions in this work. In Sec.~\ref{sec:Analytic}, using the CQSFA as a starting point, we derive analytic expressions that will be subsequently used to model ATI PADs and to describe several types of interference. In Sec.~\ref{sec:Holographic} we perform a detailed analysis of such expressions. This includes a comparison with the CQSFA and solutions of the TDSE, and an in-depth study of the physical causes of key features in several types of holographic structures, including the fan and the spider. Finally, in Sec.~\ref{sec:conclusions}, we provide the main conclusions to be taken from this work. 

\section{Background}

\label{sec:theory}

\subsection{General CQSFA expressions}

The CQSFA ATI transition amplitude from the bound state $\left\vert \psi
_{0}(t')\right\rangle =\exp [iI_{p}t']\left\vert \psi
_{0}\right\rangle $, where $I_{p}$ is the ionization potential, to a
continuum state $|\mathbf{p}_{f}(t)\rangle $ with final momentum $\mathbf{p}%
_{f}$ may be obtained from the formally exact expression
\begin{equation}
M(\mathbf{p}_{f})=-i\lim_{t\rightarrow \infty }\int_{-\infty }^{t}dt^{\prime
}\left\langle \mathbf{p}_{f}(t)|U(t,t')H_{I}(t')|\psi
_{0}(t')\right\rangle \,,  \label{eq:transitionampl}
\end{equation}%
where
\begin{equation}
U(t,t')=\mathcal{T}\exp \bigg [i\int_{t'}^{t}H(t^{\prime
})dt'\bigg],  \label{eq:timeev}
\end{equation}%
with $\mathcal{T}$ denoting time ordering, gives the time evolution operator
related to the full Hamiltonian
\begin{equation}
H(t)=H_{a}+H_{I}(t)  \label{eq:Hamiltonian}
\end{equation}%
evolving from an initial time $t'$ to a final time $t$. In Eq.~(\ref%
{eq:Hamiltonian}),
\begin{equation}
H_{a}=\frac{\hat{\mathbf{p}}^{2}}{2}+V(\hat{\mathbf{r}})
\end{equation}%
gives the field-free one-electron atomic Hamiltonian, where $\hat{\mathbf{r}}
$ and $\hat{\mathbf{p}}$ yield the position and momentum operators,
respectively, and $V(\hat{\mathbf{r}})$ is the binding potential, which is
chosen to be of Coulomb type. The Hamiltonian
\begin{equation}
H_{I}(t)=-\hat{\mathbf{r}}\cdot \mathbf{E}(t)
\end{equation}%
gives the interaction with the external field $\mathbf{E}(t)$ in the length
gauge. Note that the Strong-Field Approximation (SFA) transition amplitude
for the direct electrons, which do not undergo collisions with the core
after ionization, can be obtained by replacing the full time evolution
operator (\ref{eq:timeev}) by its Volkov counterpart, in which the influence
of the binding potential is neglected.

Employing a closure relation in the initial momentum and a path-integral
formulation from an initial velocity $\tilde{\mathbf{p}}_{0}=\mathbf{p}%
_{0}(t')+\mathbf{A}(t')$ to a final velocity $\tilde{%
\mathbf{p}}_{f}(t)=\mathbf{p}_{f}(t)+\mathbf{A}(t)$ in Eq.~(\ref%
{eq:transitionampl})\cite{Lai2015,Maxwell2017}, where $\mathbf{A}(\tau )$, $%
\tau =t,t'$ is the vector potential, one obtains
\begin{eqnarray}
M(\mathbf{p}_{f}) &=&-i\lim_{t\rightarrow \infty }\int_{-\infty
}^{t}dt'\int d\mathbf{\tilde{p}}_{0}\int_{\mathbf{\tilde{p}}_{0}}^{%
\mathbf{\tilde{p}}_{f}(t)}\mathcal{D}'\mathbf{\tilde{p}}\int \frac{%
\mathcal{D}\mathbf{r}}{(2\pi )^{3}}  \nonumber \\
&& \times e^{iS(\mathbf{\tilde{p}},\mathbf{r},t,t')}\langle \mathbf{%
\tilde{p}}_{0}|H_{I}(t')|\psi _{0}\rangle \,,  \label{eq:CQSFA}
\end{eqnarray}
where $\tilde{\mathbf{p}}$ and $\mathbf{r}$ give the intermediate
velocity and coordinate, respectively, the symbols $\mathcal{D}'%
\mathbf{p}$ and $\mathcal{D}\mathbf{r}$ denote the integration measures for the
path integrals \cite{Kleinert2009,Lai2015}, and the prime indicates a
restriction. The action is given by
\begin{equation}
S(\mathbf{\tilde{p}},\mathbf{r},t,t')=I_{p}t^{\prime
}-\int_{t'}^{t}[\dot{\mathbf{p}}(\tau )\cdot \mathbf{r}(\tau )+H(%
\mathbf{r}(\tau ),\mathbf{p}(\tau ),\tau ]d\tau ,  \label{eq:stilde}
\end{equation}%
where
\begin{equation}
H(\mathbf{r}(\tau ),\mathbf{p}(\tau ),\tau )=\frac{1}{2}\left[ \mathbf{p}%
(\tau )+\mathbf{A}(\tau )\right] ^{2}+V(\mathbf{r}(\tau ))
\label{eq:Hamiltonianpath}
\end{equation}%
is the parameterized Hamiltonian. The binding potential reads
\begin{equation}
V(\mathbf{r}(\tau ))=-\frac{C}{\sqrt{\mathbf{r}(\tau )\cdot \mathbf{r}(\tau )%
}},  \label{eq:potential}
\end{equation}%
where $0\leq C\leq 1$ is an effective coupling. One should note that Eq.~(\ref{eq:CQSFA}) excludes transitions between bound states, such as excitation or relaxation.

The
above-stated integrals are performed using the stationary phase method. We
seek solutions for $t'$, $\mathbf{r}(\tau )$ and $\mathbf{p}(\tau )$
so that the action is stationary. This leads to the saddle-point equations
\begin{equation}
\frac{\left[ \mathbf{p}(t')+\mathbf{A}(t')\right] ^{2}}{2}%
+V(\mathbf{r}(t'))=-I_{p},  \label{eq:tunnel}
\end{equation}%
\begin{equation}
\mathbf{\dot{p}}(\tau)=-\nabla _{r}V(\mathbf{r}(\tau )),  \label{eq:p-spe}
\end{equation}%
and
\begin{equation}
\mathbf{\dot{r}}(\tau)=\mathbf{p}(\tau)+\mathbf{A}(\tau ).  \label{eq:q-spe}
\end{equation}%
Eq.~(\ref{eq:tunnel}) describes the conservation of energy upon tunnel
ionization, and the remaining equations give the electron motion in the
continuum. We employ a two-pronged contour \cite%
{Popruzhenko2008JMO,Yan2012,Torlina2012PRA,Torlina2013}, whose first arm is
parallel to the imaginary-time axis, from $t'=t_{r}^{\prime
}+it_{i}'$ to $t_{r}'$, and whose second arm is taken to
be along the real time axis, from $t_{r}'$ to $t$, respectively.
This yields
\begin{equation}
S(\mathbf{\tilde{p}},\mathbf{r},t,t')=S^{\mathrm{tun}}(\mathbf{%
\tilde{p}},\mathbf{r},t_{r}',t')+S^{\mathrm{prop}}(\mathbf{%
\tilde{p}},\mathbf{r},t,t_{r}'),
\end{equation}%
where $S^{\mathrm{tun}}(\mathbf{\tilde{p}},\mathbf{r},t_{r}^{\prime
},t')$ and $S^{\mathrm{prop}}(\mathbf{\tilde{p}},\mathbf{r}%
,t,t_{r}')$ give the action along the first and second part of the
contour, respectively. These expressions are associated with the electron's
tunnel ionization and continuum propagation, respectively.

We assume the electron momentum to be approximately constant during tunnel
ionization, so that
\begin{eqnarray}
S^{\mathrm{tun}}(\mathbf{\tilde{p}},\mathbf{r},t_{r}',t')
&=&I_{p}(it_{i}')-\frac{1}{2}\int_{t'}^{t_{r}'}%
\left[ \mathbf{p}_{0}+\mathbf{A}(\tau )\right] ^{2}d\tau  \nonumber \\
&&-\int_{t'}^{t_{r}'}V(\mathbf{r}_{0}(\tau ))d\tau ,
\label{eq:stunn}
\end{eqnarray}%
where the tunnel trajectory $\mathbf{r}_{0}$ is defined by%
\begin{equation}
\mathbf{r}_{0}(\tau )=\int_{t'}^{\tau }(\mathbf{p}_{0}+\mathbf{A}%
(\tau '))\mathrm{d}\tau '.  \label{eq:tunneltraj}
\end{equation}%
Consequently, the binding potential may be neglected in Eq.~(\ref{eq:tunnel}%
), which becomes
\begin{equation}
\frac{1}{2}\left[ \mathbf{p}_{0}+\mathbf{A}(t')\right] ^{2}+I_{p}=0.
\label{eq:tp-spe}
\end{equation}%
The contour for $S^{\mathrm{tun}}(\mathbf{\tilde{p}},\mathbf{r}%
,t_{r}',t')$ is calculated from the origin up to the
tunnel exit, which is
\begin{equation}
z_{0}=\mathrm{Re}[r_{0\parallel}(t'_r)],
\label{eq:tunnelexit}
\end{equation}%
where $r_{0\parallel}$ is the component of the tunnel trajectory parallel to the laser-field polarization \cite{PPT1967}. The
action related to the continuum propagation reads
\begin{eqnarray}
S^{\mathrm{prop}}(\mathbf{\tilde{p}},\mathbf{r},t,t_{r}')
&=&I_{p}t_{r}'-\frac{1}{2}\int_{t_{r}'}^{t}\left[ \mathbf{p%
}(\tau )+\mathbf{A}(\tau )\right] ^{2}d\tau  \nonumber \\
&&-2\int_{t_{r}'}^{t}V(\mathbf{r}(\tau ))d\tau ,  \label{eq:sprop}
\end{eqnarray}%
where the factor 2 multiplying the binding potential comes from
\begin{equation}
\mathbf{r}\cdot \dot{\mathbf{p}}=-\mathbf{r}\cdot \nabla _{r}V(r)=V(r),
\label{eq:virial}
\end{equation}%
as discussed in \cite{Shvetsov-ShilovskiPRA2016,Maxwell2017}, and 
\begin{equation}
\mathbf{r}(\tau )=\int_{t_r'}^{\tau }(\mathbf{p}(\tau ')+\mathbf{A}%
(\tau '))\mathrm{d}\tau ' \label{eq:conttraj}
\end{equation}
gives the spatial coordinate related to the continuum propagation. 

The potential is included in Eqs.~(\ref{eq:q-spe}) and (\ref{eq:p-spe}) and
in the action (\ref{eq:sprop}), which are solved for a specific final
momentum $\mathbf{p}_{f}$, and for $t\rightarrow \infty $. \ The solutions
from the first and the second arm of the contour are then matched at the
tunnel exit.

Within the saddle-point approximation, the Coulomb corrected transition
amplitude reads as
\begin{equation}
M(\mathbf{p}_{f}) \propto -i\lim_{t\rightarrow \infty }\sum_{s}\bigg\{\det %
\bigg[\frac{\partial \mathbf{p}_{s}(t)}{\partial \mathbf{r}_{s}(t_{s})}\bigg]%
\bigg\}^{-1/2}  \mathcal{C}(t_{s})e^{iS(\mathbf{\tilde{p}}_{s},\mathbf{r}%
_{s},t,t_{s}))},\label{eq:MpPathSaddle}
\end{equation}%
where $t_{s}$, $\mathbf{p}_{s}$ and $\mathbf{r}_{s}$ are given by Eqs.~(\ref%
{eq:q-spe})-(\ref{eq:tp-spe}) and
\begin{equation}
\mathcal{C}(t_{s})=\sqrt{\frac{2\pi i}{\partial ^{2}S(\mathbf{\tilde{p}}_{s},\mathbf{r}%
		_{s},t,t_{s})%
/\partial t_{s}^{2}}}\langle \mathbf{p}+\mathbf{A}(t_{s})|H_{I}(t_{s})|\Psi
_{0}\rangle .  \label{eq:Prefactor}
\end{equation}%
In practice, we use the stability factor $\partial \mathbf{p}%
_{s}(t)/\partial \mathbf{p}_{s}(t_{s})$ instead of that given in Eq.~(\ref{eq:MpPathSaddle}), which may be obtained applying a Legendre transformation. The action
will not be modified under this transformation as long as the electron
starts from the origin. We normalize Eq.~(\ref{eq:MpPathSaddle}) so that in
the limit of vanishing binding potential the SFA is recovered. Throughout, we refer to the product of Eq.~(\ref{eq:Prefactor}) with the stability factor as ``the prefactors". 

\subsection{Model}

For simplicity, in the results that follow, we will consider a linearly
polarized monochromatic field
\begin{equation}
\mathbf{E}(t)=E_0 \sin(\omega t)\hat{e}_{\parallel}.  \label{eq:efield}
\end{equation}
This corresponds to the vector potential
\begin{equation}
\mathbf{A}(t)=2 \sqrt{U_p} \cos(\omega t)\hat{e}_{\parallel},
\label{eq:afield}
\end{equation}
where $\hat{e}_{\parallel}$ gives the unit vector in the direction of the
driving-field polarization and $U_p$ is the ponderomotive energy.

This leads to the action
\begin{eqnarray}
&\hspace*{-1mm}S^{\mathrm{tun}}(\mathbf{\tilde{p}},\mathbf{r},t_r^{\prime
},t')= i\left(\ip+\frac{1}{2}\mathbf{p}_0^2 +
\up \right) t'_i -\int^{t'_r}_{t'} V(\mathbf{r}_0(\tau))\mathrm{d}\tau  \nonumber \\
& \hspace{1.8cm} + 2\frac{\sqrt{\up}p_{0 \parallel}}{\omega} \left[
\sin(\omega t')-\sin(\omega t'_r)\right]  \nonumber \\
& \hspace{1.8cm} +\frac{\up}{2\omega}\left[\sin(2\omega
t') -\sin(2\omega t'_r) \right]  \label{eq:stunn_ex}
\end{eqnarray}
and
\begin{eqnarray}
&\hspace{-3mm}S^{\mathrm{prop}}(\mathbf{\tilde{p}},\mathbf{r},t,t^{\prime
}_r)= \left(\ip + \frac{1}{2}\mathbf{p}_f^2 + \mathrm{U}_%
\mathrm{p} \right) t'_r + \frac{2\sqrt{\up}%
p_{f\parallel}}{\omega}\sin(\omega t'_r)  \nonumber \\
&\hspace{13mm} +\frac{\up}{2\omega}\sin(2\omega t^{\prime
}_r) -\frac{1}{2}\int_{t'_r}^{t}\pmb{\mathcal{P}}(\tau)\cdot (%
\pmb{\mathcal{P}}(\tau)+2\mathbf{p}_f+2\mathbf{A}(\tau))\mathrm{d}\tau  \nonumber \\
&\hspace{13mm}  -2
\int^{t'}_{t'_r} V(\mathbf{r}(\tau))\mathrm{d}\tau,
\label{eq:sprop_ex}
\end{eqnarray}
in the first and second part of the contour, respectively, where $%
p_{j\parallel}$, with $j=0,f$, yield the electron momentum components
parallel to the laser-field polarization and $\mathbf{p}(\tau)=%
\pmb{\mathcal{P}}(\tau)+\mathbf{p}_f$. Eqs.~(\ref{eq:stunn_ex}) and (\ref%
{eq:sprop_ex}) can be combined in such a way that
\begin{eqnarray}
S(\mathbf{\tilde{p}},\mathbf{r},t,t') &=\left(\ip+ \up \right) t'
+\frac{1}{2}\mathbf{p}^2_f t'_{r}
+\frac{i}{2}\mathbf{p}^2_{0} t'_{i}
+\frac{\up}{2\omega}\sin(2\omega t')  \nonumber \\
& +\frac{2\sqrt{\up}}{\omega}\left[
p_{0\parallel}\sin(\omega t')-(p_{0\parallel}-p_{f\parallel})\sin(\omega t'_{r})
\right]
-\int^{t'_{r}}_{t'}  V(\mathbf{r}_{0}(\tau))\mathrm{d}\tau  \nonumber \\
& -\frac{1}{2}\int_{t'_{r}}^{t}\pmb{\mathcal{P}}(\tau)\cdot (\pmb{\mathcal{P}}(\tau)+2\mathbf{p}_f+2\mathbf{A}(\tau))\mathrm{d}\tau
-2 \int^{t}_{t'_{r}} V(\mathbf{r}(\tau))
\mathrm{d}\tau.  \label{eq:s_ex}
\end{eqnarray}
In the limit $C\rightarrow 0$,  $V(\mathbf{r})\rightarrow 0$, $\mathbf{p}%
_f\rightarrow \mathbf{p}_0\rightarrow\mathbf{p}$ and $\pmb{\mathcal{P}}%
\rightarrow 0$, so that the standard SFA action for direct ATI electrons is recovered.

The ionization time $t'_{ec}$ associated with an event $e$ occurring in a cycle
$c$ is obtained analytically by solving the saddle-point equation (\ref%
{eq:tp-spe}). For clarity, we will employ this notation instead of the index
$s$ used in the transition amplitude (\ref%
{eq:MpPathSaddle}). This yields
\begin{equation}
t'_{ec}=\frac{2\pi n}{\omega }\pm \frac{1}{\omega }\arccos \left( \frac{%
-p_{0\parallel }\mp i\sqrt{2I_{\mathrm{p}}+p_{0\perp }^{2}}}{2\sqrt{U_{%
\mathrm{p}}}}\right) ,
\end{equation}
where $p_{0\perp }$ denotes the component of the initial momentum
perpendicular to the laser-field polarization. 
Within a field cycle, each event $e$ may be associated with a specific type of orbit. In this work, we employ the
same orbit classification as in \cite{Lai2015,Lai2017,Maxwell2017}, which
has been first introduced in \cite{Yan2010}, where there are four types of orbits for a given final momentum.
For orbit 1, the electron reaches the detector directly without undergoing a
deflection. Consequently, the final and initial momenta point in the same
direction. For orbits 2 and 3, the electron is freed on the opposite side,
so that, in order to reach the detector, it must change its parallel
momentum component. The main difference is that, for orbit 2, the
perpendicular momentum component does not change sign, while for orbit 3 it
does. Finally, if the electron is freed along orbit 4, it leaves from the same
side of the detector, but its transverse momentum component changes.
Physically, this means that the electron goes around the core before
reaching the detector at a later time. Our previous work indicates that the transition amplitude associated with orbit 4 is much
smaller than those related to the remaining orbits. Hence, the contribution
of this orbit can be neglected. If the Coulomb potential is neglected in the
continuum, which is the case for the strong-field approximation (SFA), only
orbits type 1 and 2 exist. The latter exhibits a degeneracy that is lifted
in the presence of the Coulomb potential, leading to orbits 2 and 3 \cite%
{Lai2015}.

The specific solutions for orbits 1, 2 and 3 within a particular field cycle $%
c $ and parallel momentum component $p_{0\parallel }>0$ are
\begin{eqnarray}
t'_{1c} &=&\frac{1}{\omega }\arccos \left( \frac{-p_{0\parallel }-i\sqrt{2I_{%
\mathrm{p}}+p_{0\perp }^{2}}}{2\sqrt{U_{\mathrm{p}}}}\right)  \label{eq:t1s}
\\
t'_{ec} &=&\frac{2\pi }{\omega }-\frac{1}{\omega }\arccos \left( \frac{%
-p_{0\parallel }+i\sqrt{2I_{\mathrm{p}}+p_{0\perp }^{2}}}{2\sqrt{U_{\mathrm{p%
}}}}\right) ,  \label{eq:t2s}
\end{eqnarray}%
with $e=2,3$. One should note that $t_{2c}$ and $t_{3c}$ differ, due to the
distinct initial momenta for orbits 2 and 3. For $p_{0\parallel }<0$, the
situation reverses, so that $t_{1c}$ is given by Eq.~(\ref{eq:t2s}) and the
remaining times by Eq.~(\ref{eq:t1s}). In the implementation of our model,
we solve the inverse problem, i.e., for a given final momentum $\mathbf{p}%
_{f}$, we seek the matching initial momentum $\mathbf{p}_{0}.$ This approach
has the advantage of only requiring a few contributed trajectories, each of
which is associated to the trajectory types $1$ to $3.$ In contrast, solving
the direct problem requires around $10^{8}-10^{9}$ trajectories for clear
interference patterns (for discussions of both types of implementation see,
e.g., \cite{Yan2010,Yan2012,Lai2015,Lai2017,Maxwell2017}). 
Our previous publications show that the fan- and the spider-like patterns result from the intra-cycle interference of orbits 1 and 2,  and of orbits 2 and 3, respectively \cite{Lai2017,Maxwell2017}. An additional restriction required for the fan-shaped structure to form is that the difference between the real parts of $t_{1c}$ and $t_{2c}$ should not exceed half a cycle in absolute value. Relaxing this restriction will lead to other interference structures that are commonly overlooked (for details see \cite{Maxwell2017}). 

The dominant contributors to the overall shape of the electron-momentum distributions and to the interference patterns are the imaginary and real parts of the action, respectively. The imaginary part of the action is directly related to the tunneling probability density, and it is a good indicator of the width of the barrier. Specifically for Eq.~(\ref{eq:s_ex}), $\mathrm{Im}[S]$ reads 
	\begin{eqnarray}
	S_e^{\mathrm{Im}}(t',\mathbf{p},\mathbf{r})&= \left(I_p+U_\mathrm{p}+\frac{1}{2}\mathbf{p}_{e0}^2\right)t'_{i}
	+\frac{2 p_{e0\parallel} \sqrt{U_\mathrm{p}} \cos (\omega t'_{r} )\sinh (\omega t'_{i} )}{\omega } \nonumber\\
	&	+\frac{U_\mathrm{p}\cos (2  \omega t'_{r} )\sinh (2 \omega t'_{i}) }{2 \omega }
	-\int^{t'_{r}}_{t'} \Im[V(\mathbf{r}_{e0}(\tau))]\mathrm{d}\tau,
	\label{eq:ImCQSFAAction}
	\end{eqnarray}
where $e=1,2,3$. 
The real parts give the phase differences between different types of trajectories. For the action (\ref{eq:s_ex}) and a specific 
	orbit $e$, $ \mathrm{Re}[S]$ is given by {\ {\small \medmuskip=0 mu \thinmuskip=0mu \thickmuskip%
			=0mu
			\begin{eqnarray}
			\hspace{-3mm}S_{e}^{\Re } &=&\left( I_{p}+U_{p}+\frac{1}{2}\mathbf{p}%
			_{f}^{2}\right) t'_{er}+\frac{2\sqrt{Up}}{\omega }\left( 2p_{e0\parallel}\sinh \left(
			\frac{\omega t'_{ei}}{2}\right) ^{2}+p_{zf}\right) \sin (\omega t'_{er})
			\nonumber \\
			&&+\frac{U_{p}}{2\omega }\Re \left[ \sin (2\omega t'_{e})\right] -\frac{1}{2}%
			\int_{t'_{er}}^{t}f_{e}(\tau )d\tau -\int_{t'_{e}}^{t'_{er}}\Re \left[
			V(\mathbf{r}_{e0}(\tau ))\right] d\tau   \nonumber \\
			&&-2\int_{t'_{er}}^{t}V(\mathbf{r}_{e}(\tau ))d\tau,\label{eq:sreal}
			\end{eqnarray}%
		}}where $f_{e}(\tau )=\pmb{\mathcal{P}}_{e}(\tau )\cdot (\pmb{\mathcal{P}}%
		_{e}(\tau )+2\mathbf{p}_{f}+2\mathbf{A}(\tau ))$ and $\mathbf{p}_{e}(\tau )=%
		\pmb{\mathcal{P}}_{e}(\tau )+\mathbf{p}_{f}$.
Prefactors will introduce additional biases, which do influence the shape of the PADS. They will however play a secondary role in quantum-interference effects as they vary much more slowly than the action. 

If the prefactors are neglected, one may write the ATI photoelectron probability density for $N_{c}$ cycles of the driving field and a number $n_{e}$
of relevant events per cycle as
\begin{equation}
\Omega (\mathbf{p}_{f})=\left\vert \sum_{e=1}^{n_{e}}\sum_{c=1}^{N_{c}}\exp
[iS_{ec}]\right\vert ^{2},  \label{eq:generalinterf}
\end{equation}%
where $S_{ec}$ is the action related to the $e^{th}$ event in the $c^{th}$
cycle. If the field is monochromatic, the intra and intercycle contributions
to the interference pattern are factorizable and may be reduced to
\begin{equation}
\Omega (\mathbf{p}_{f})=\Omega _{n_{e}}(\mathbf{p}_{f})\Omega _{N_{c}}(%
\mathbf{p}_{f}),  \label{eq: generalizedinterf}
\end{equation}%
where
\begin{equation}
\Omega _{N_{c}}(\mathbf{p}_{f})=\frac{\cos \left[ \frac{2\pi N_{c}}{\omega }%
	\left( \mathrm{I}_{\mathrm{p}}+\mathrm{U}_{\mathrm{p}}+\frac{1}{2}\mathbf{p}%
	_{f}^{2}\right) \right] -1}{\cos \left[ \frac{2\pi }{\omega }\left( \mathrm{%
		I}_{\mathrm{p}}+\mathrm{U}_{\mathrm{p}}+\frac{1}{2}\mathbf{p}_{f}^{2}\right) %
	\right] -1}  \label{eq:nintercycle}
\end{equation}%
gives the inter-cycle rings and $\Omega _{n_{e}}(\mathbf{p}_{f})$ is
associated with intra-cycle interference. In $\Omega_{n_e} (\mathbf{p}_{f})$, we will consider \textit{pairs} of orbits within a cycle, as these are sufficient for describing holographic structures.  For detailed
derivations and further discussions see our publication \cite{Maxwell2017}. 
\section{Analytic expressions}
\label{sec:Analytic}
We will now provide analytic approximations for the sub-barrier dynamics and the continuum propagation. In order to make Eq.~(\ref{eq:sreal}) analytically solvable, we employ the low-frequency approximation and several simplifying assumptions upon the final and intermediate momenta. The low-frequency approximation has been used in \cite{Yan2012} to derive sub-barrier corrections, and in our previous publication \cite{Maxwell2017} for computing analytical single-orbit probability distributions from Eq.~(\ref{eq:ImCQSFAAction}).

The quantities of interest are the under the barrier potential integral
\begin{equation}
\mathcal{I}_{V_T}=-\int^{t'_{er}}_{t'_e} V(\mathbf{r}_{e0}(\tau))\mathrm{d}\tau,
\label{eq: Ivt}
\end{equation}
the potential integral 
\begin{equation}
\mathcal{I}_{V_C}= -\int_{t'_{er}}^{t}V(\mathbf{r}_{e}(\tau ))d\tau
\label{eq:IVc}
\end{equation}
related to the continuum propagation, and  the phase difference
\begin{equation}
\mathcal{I}_{\mathcal{P}_e}=-\frac{1}{2}%
\int_{t'_{er}}^{t}\pmb{\mathcal{P}}_{e}(\tau )\cdot (\pmb{\mathcal{P}}%
_{e}(\tau )+2\mathbf{p}_{f}+2\mathbf{A}(\tau ))d\tau
\label{eq:Imom}
\end{equation}
 due to the electron's final and intermediate momentum being different, as it is accelerated by the residual binding potential. 

\subsection{The under-the-barrier integral and single-orbit distributions}

An analytic expression for Eq.~(\ref{eq: Ivt}) has been computed in \cite{Maxwell2017} and reads
	\begin{equation}
	\hspace*{-2.5cm}\int^{t'_{r}}_{t'-i\Delta \tau_i}\hspace*{-0.5cm} V(\mathbf{r}_0(\tau))\mathrm{d}\tau= i \ln \left[\left(\frac{t'_i \left(\chi \eta(t'_i-\Delta \tau_i) -p^2_{0\perp}+\sqrt{-\mathbf{p}^2_{0\perp}+\eta(t'_i-\Delta \tau_i)^2}\sqrt{-\mathbf{p}^2_{0\perp}+\chi^2}\right)}{\Delta \tau_i \left(\chi \eta(0) -p^2_{0\perp}+\sqrt{-\mathbf{p}^2_{0\perp}+\eta(0)^2}\sqrt{-\mathbf{p}^2_{0\perp}+\chi^2}\right)}\right)^{C/\sqrt{-\mathbf{p}^2_{0\perp}+\chi^2}}\right],
	\label{eq:Vintanalytic}
	\end{equation}
	where
	\begin{equation}
	\chi=i ({p}_{0\perp}+A(t'_{r}))-t'_{i} \dot{A}(t'_{r}) 
	\end{equation}
	\begin{equation}
	\eta(\tau_i)=i ({p}_{0\parallel}+{A}(t'_{r}))-\frac{1}{2}(t'_{i}+\tau_i) \dot{A}(t'_{r}),
	\end{equation}
	$0\le C \le 1$ is the effective Coulomb coupling and the subscripts $e$ have been dropped for simplicity.    One should note that the lower integration limit has been slightly modified in order to avoid a logarithmic divergence, and that $\Delta \tau_i$ can be chosen to be arbitrarily small. This divergence can be removed by a regularization procedure.  Eq.~(\ref{eq:Vintanalytic}) can be split into a non-divergent and a divergent part, which can be treated separately. This gives
	\begin{equation}
	\mathcal{I}_{V_T}= \widetilde{\mathcal{I}}_{V_T}+\mathcal{I}_{\mathrm{div}},
\end{equation}
with
	\begin{equation}
	\hspace{-2cm}
\widetilde{\mathcal{I}}_{V_T}=  i \ln \left[\left(\frac{t'_{i} \left(\chi \eta(t'_{i}-\Delta \tau_i) -p^2_{0\perp}+\sqrt{-\mathbf{p}^2_{0\perp}+\eta(t'_{i}-\Delta \tau_i)^2}\sqrt{-\mathbf{p}^2_{0\perp}+\chi^2}\right)}
	{\chi \eta(0) \nonumber -p^2_{0\perp}+\sqrt{-\mathbf{p}^2_{0\perp}+\eta(0)^2}\sqrt{-\mathbf{p}^2_{0\perp}+\chi^2}}\right)^{C/\sqrt{-\mathbf{p}^2_{0\perp}+\chi^2}}\right]
		\label{eq:NonDiv}
	\end{equation} 
	and
	\begin{equation}
	\mathcal{I}_{\mathrm{div}}=
-i C/\sqrt{-\mathbf{p}^2_{0\perp}+\chi^2} \ln(\Delta \tau_i).
	\label{eq:Div}
	\end{equation}
	In Eq.~(\ref{eq:NonDiv}),  $\Delta \tau_i\rightarrow0$ leads to $\eta(t'_{i}-\Delta \tau_i)\rightarrow\chi$, while Eq.~(\ref{eq:Div}), when added into the action, will act like a prefactor. Explicitly, 
	\begin{eqnarray}
	\exp(i\widetilde{\mathcal{I}}_{V_T} +i\mathcal{I}_{\mathrm{div}})&=&\exp(i\widetilde{\mathcal{I}}_{V_T})\exp(C/\sqrt{-\mathbf{p}^2_{0\perp}+\chi^2} \ln(\Delta \tau_i))\\ \nonumber
	&=&\Delta \tau_i^{C/\sqrt{-\mathbf{p}^2_{0\perp}+\chi^2}}\exp(i\widetilde{\mathcal{I}}_{V_T})
	\end{eqnarray}
	To remove this factor from the expression we can use the freedom that we may tend $\Delta \tau_i$ to zero via any route. We can set $\Delta \tau_i=\delta^{\sqrt{-\mathbf{p}^2_{0\perp}+\chi^2}/C}$, where $\delta$ is a parameter that can used for all orbits to tend  $\Delta \tau_i$ to zero. This will lead to a common factor $\delta$, which will affect the overall yield but not the interference patterns. Hence, it can be removed. 
	
	The regularized expression for Eq.~(\ref{eq:Vintanalytic}) then reads
	\begin{equation}
	\hspace*{-2cm}
	\mathcal{I}_{V_T}= i {C/\sqrt{-\mathbf{p}^2_{0\perp}+\chi^2}}\ln \left[\underbrace{\left(\frac{2t'_i \left(\chi^2 -p^2_{0\perp}\right)}
		{\chi \eta(0) -p^2_{0\perp}+\sqrt{-\mathbf{p}^2_{0\perp}+\eta(0)^2}\sqrt{-\mathbf{p}^2_{0\perp}+\chi^2}}\right)}_{\mathcal{F}}\right].
	\label{eq:ITunFinal}
	\end{equation}	
	
\begin{figure}
	\includegraphics[width=\textwidth]{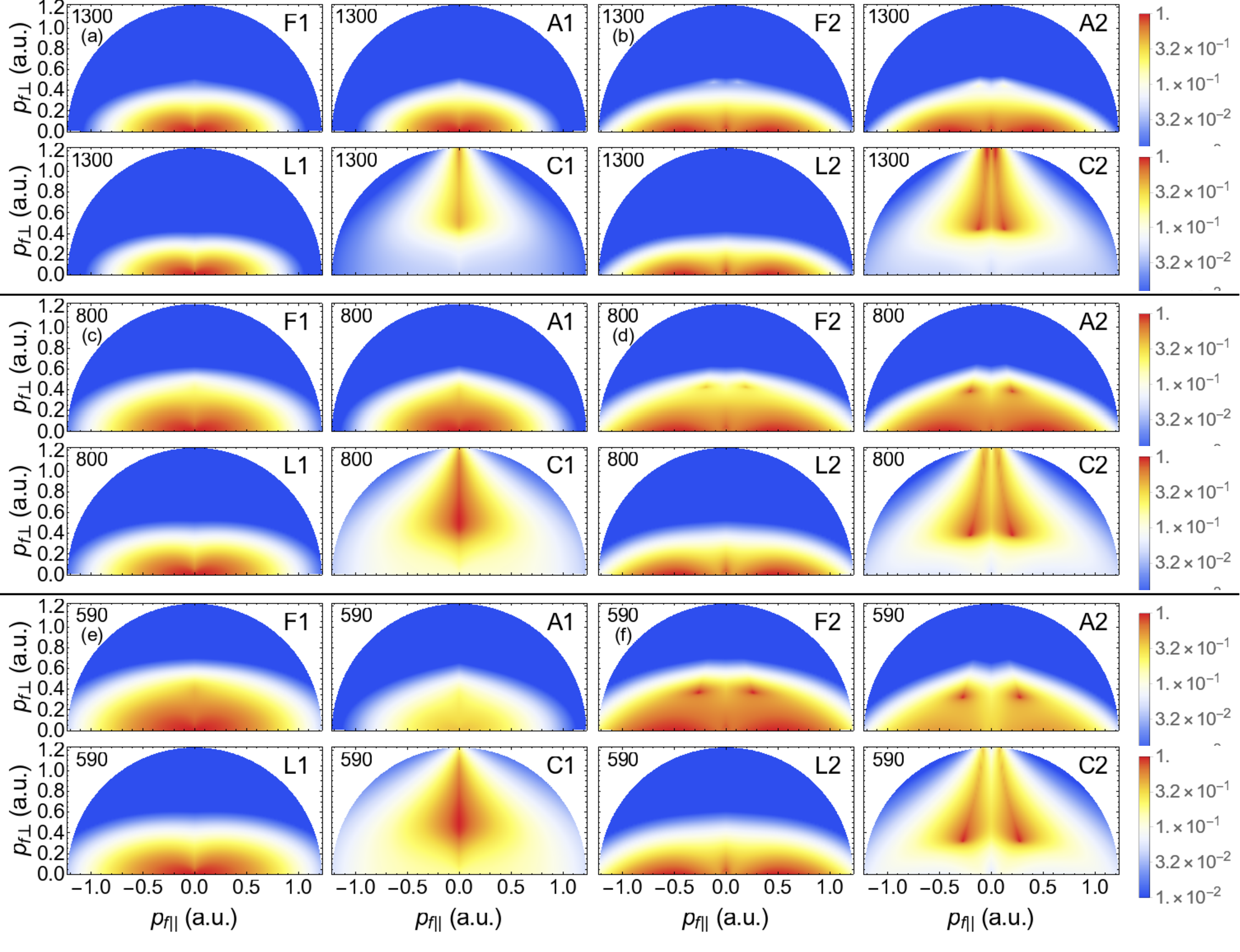}
	\caption{Single-orbit ATI photoelectron angle-resolved distributions (PADs) computed without prefactors for hydrogen ($I_p=0.5$ a.u.) and orbits 1 and 2. The lower case letters on the top left corner correspond to the field parameters  $(I,\lambda)=(7.5\times 10^{13} \mathrm{W/cm}^2,1300 \hspace*{0.1cm}\mathrm{nm})$ [panels (a) and (b)], $(I,\lambda)=(2.0\times 10^{14} \mathrm{W/cm}^2,800\hspace*{0.1cm} \mathrm{nm})$ [panels (c) and (d)] and $(I,\lambda)=(3.75\times 10^{14} \mathrm{W/cm}^2,590\hspace*{0.1cm} \mathrm{nm})$ [panels (e) and (f)], where $I$ and $\lambda$ give the field intensity and wavelength, respectively. This yields a Keldysh parameter $\gamma \approx  0.75$. The acronyms $\mathrm{F}n$ $(n=1,2)$, $\mathrm{A}n$ $(n=1,2)$ on the right top corner indicate the full and analytic CQSFA solution for orbits 1 or 2, while $\mathrm{L}n$ $(n=1,2)$ and $\mathrm{C}n$ $(n=1,2)$ give the laser and Coulomb terms of the analytic expressions as defined in Eqs.~(\ref{eq:analyticsingle}) and (\ref{eq:analyticsing2}). The numbers on the top left corner of each panel give the driving-field wavelengths.  The density plots have been represented in a logarithmic scale and normalized to the highest yield in each panel. The thick horizontal lines separate panels with different field parameters.}
	\label{fig:SingleOrbNoPref}
\end{figure}

In Fig.~\ref{fig:SingleOrbNoPref}, we plot single-orbit distributions computed for orbits 1 and 2 using the full CQSFA, and the analytic approximation given by Eq.~(\ref{eq:ITunFinal}). 
In order to facilitate a comparison, the prefactors have not been included. Overall, there is little discrepancy between the analytical approximation and the full CQSFA. This is because the single-orbit plots will vary only with the imaginary part of the action, which occurs exclusively along the tunnel trajectory. The momentum along the tunnel trajectory is already taken to be constant. Thus, the only difference between both models is the long wavelength approximation used to integrate the potential. This additional approximation is quite accurate along the tunnel trajectory, partly because the path along the imaginary time axis is relatively short, typically  well under half a cycle. Furthermore, the trigonometric functions turn into hyperbolic functions, which are easily approximated. 
For both orbits, there is a double peaked structure in the analytic and full CQSFA solutions, and the yield becomes suppressed at the origin (see upper panels in Figs.~\ref{fig:SingleOrbNoPref}(a) to (f)).  This indicates that, in the presence of the Coulomb potential, the electron must have a non-vanishing momentum to reach the continuum with a high probability \cite{Maxwell2017}. For orbit 2, this structure is particularly visible and spreads to a larger momentum region as the driving-field frequency increases (see upper rows in Figs.~\ref{fig:SingleOrbNoPref}(b), (d) and (f)). Both the analytic and full CQSFA exhibit sharply focused spots in the PADs computed with orbit 2, which become more prominent as the laser frequency increases. The analytic expressions overestimate these spots. This can be seen by comparing panels $\mathrm{F}2$ and $\mathrm{A}2$ in Figs.~\ref{fig:SingleOrbNoPref}(b), (d) and (f).

%

%

Using the analytic model we can break down these effects to find their origin. The single-orbit distributions are entirely governed by the imaginary part of the action, which can be written as
\begin{eqnarray}
	S^{\Im}(\mathbf{p},\mathbf{r},t)&=&\frac{\up}{\omega}\mathcal{G}(\xi)
	-i\Re\left[\frac{C}{\sqrt{-\mathbf{p}_{0\perp}^2+\chi^2}}\ln\left(\mathcal{F}\right)\right],\\
	\mathcal{G}(\xi)&=&(1+2|\xi|^2)\arccos(\xi)-4 \xi_r\Im[\sqrt{1-\xi^2}]\nonumber \\
	&+&\Im[\xi \sqrt{1-\xi^2}]
	\label{eq:analyticsingle}
\end{eqnarray}where $\mathcal{F}$ is equal to the argument of the logarithm in Eq.~(\ref{eq:ITunFinal}), related to the binding potential, $\mathcal{G}(\xi)$ is the unit-less SFA-like part of the action, associated with the laser-induced dynamics, and  $\xi=\cos(\omega t_e')$ is a unit-less variables that has been used to replace initial momentum and time. 
%
We can separate these parts when we consider the single-orbit probability distribution, so that
\begin{eqnarray}
	|\exp(i S(\mathbf{p},\mathbf{r},t))|^2&=&\exp(-2S^{\Im}(\mathbf{p},\mathbf{r},t))\nonumber \\
	&=& \left|\mathcal{F}^{\frac{C}{\sqrt{-\mathbf{p}_{0\perp}^2+\chi^2}}}\right|^2
	\exp\left(-\frac{2\up}{\omega}\mathcal{G}(\xi)\right).
	\label{eq:analyticsing2}
\end{eqnarray}
The SFA-like part has a clear $\omega$ dependence and  contributes the most to the final shape, hence the apparent overall scaling with $\omega$ seen in the upper parts of Figs.~\ref{fig:SingleOrbNoPref}(a)--(f) [panels $\mathrm{A}n$ and $\mathrm{F}n$ $(n=1,2)$]. The potential-dependent prefactor scales in a non-trivial way.  The figure also shows that the  contributions from the SFA-like terms and the potential integrals $\mathcal{I}_{V_T}$, plotted in panels $\mathrm{L}n$ and $\mathrm{C}n$ $(n=1,2)$, respectively, mostly occupy different momentum regions. The SFA-like part of the action is more located near the $p_{\parallel}$ axis, while the Coulomb contribution leads to an elongated structure near the $p_{\perp}$ axis. For orbit 1, this structure is single peaked, but for orbit 2 it exhibits a clear suppression at $p_{\parallel}=0$, with two distinct peaks around this axis. There is also a lower momentum bound for this structure, which decreases for higher frequencies.  This will increase the overlap between the Coulomb and laser-field contributions for a shorter wavelength. This means that features such as the two spots in orbit 2, most visible for 590 nm, are due to an increasing overlap of these two parts. 

\begin{figure}
	\includegraphics[width=\textwidth]{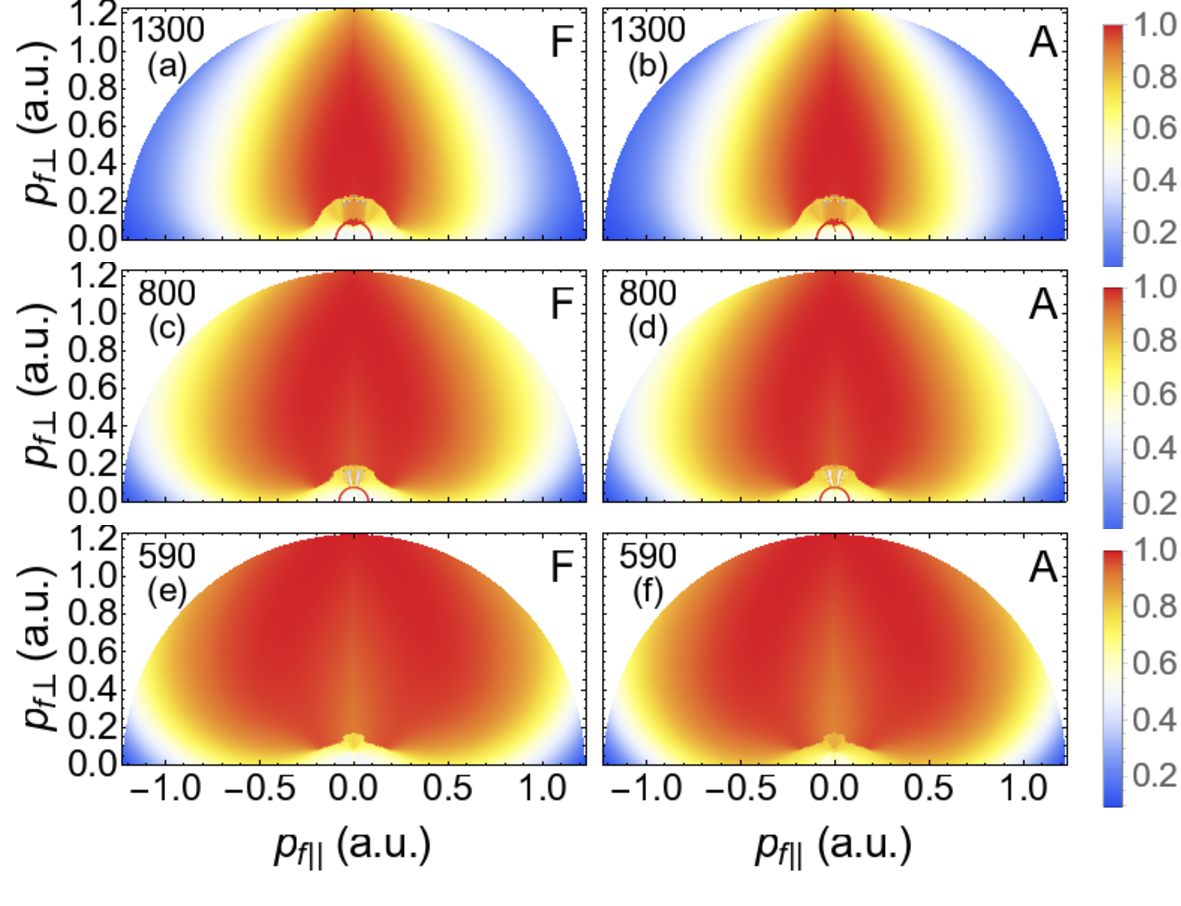}
	\caption{Single-orbit ATI PADs computed without prefactors for hydrogen ($I_p=0.5$ a.u.) and orbit 3. The top, middle and bottom panels have been calculated for the field parameters  $(I,\lambda)=(7.5\times 10^{13} \mathrm{W/cm}^2,1300 \hspace*{0.1cm}\mathrm{nm})$ [panels (a) and (b)], $(I,\lambda)=(2.0\times 10^{14} \mathrm{W/cm}^2,800\hspace*{0.1cm} \mathrm{nm})$ [panels (c) and (d)] and $(I,\lambda)=(3.75\times 10^{14} \mathrm{W/cm}^2,590\hspace*{0.1cm} \mathrm{nm})$ [panels (e) and (f)], where $I$ and $\lambda$ give the field intensity and wavelength, respectively. The upper-case letters F and A on the right top corner of each panel indicate the full and analytic CQSFA solutions and the numbers on the top left corners give the driving-field wavelength.  The density plots have been represented in a linear scale and normalized to the highest yield in each panel. }
	\label{fig:Orb3NoPref}
\end{figure}
\begin{figure}
	\includegraphics[width=\textwidth]{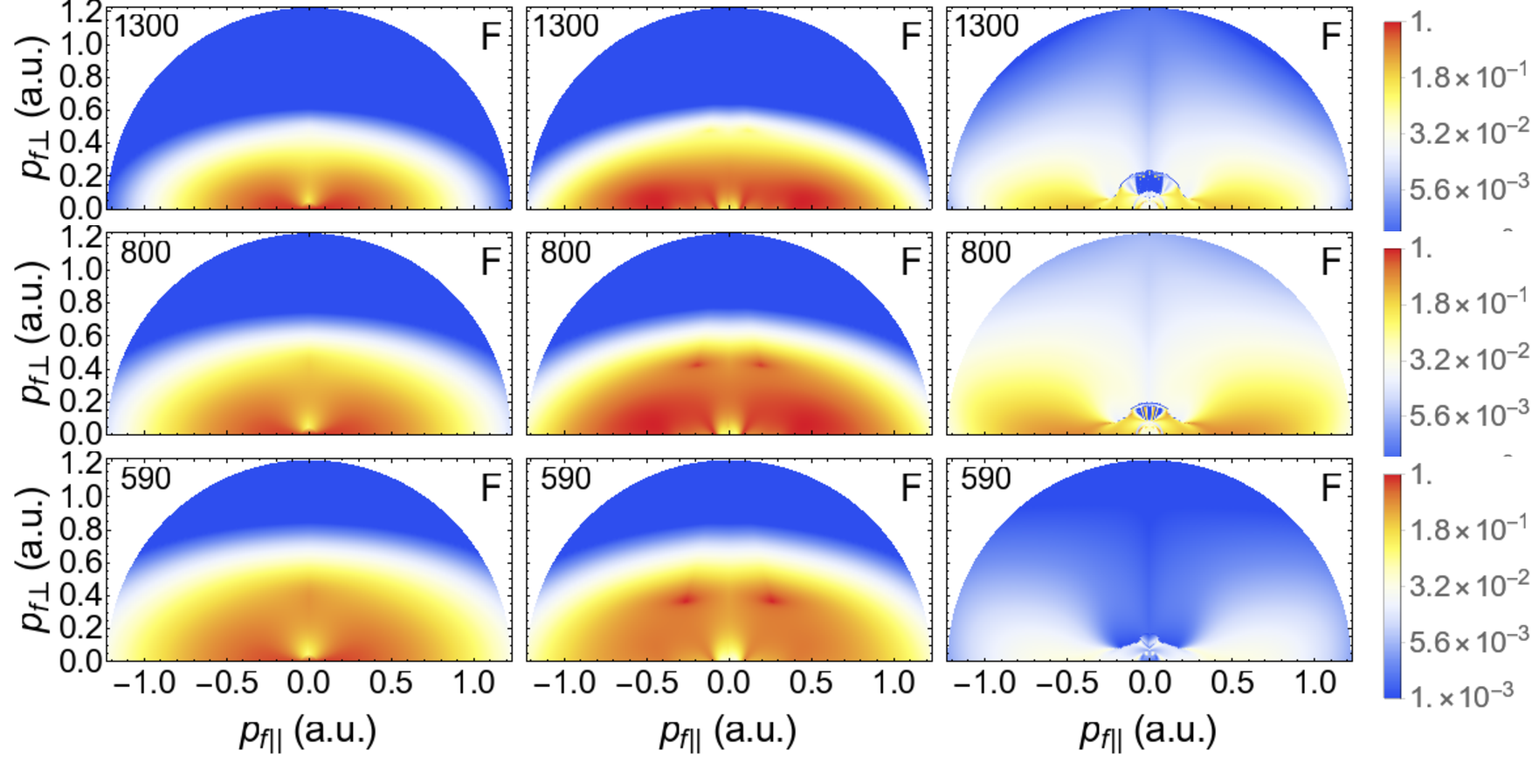}
	\caption{Single-orbit PADs for Hydrogen ($I_p=0.5$ a.u.), including all prefactors and using the full CQSFA, for orbits 1, 2 and 3 (left, middle and right columns, respectively). The top, middle and bottom panel have been calculated for the field parameters  $(I,\lambda)=(7.5\times 10^{13} \mathrm{W/cm}^2,1300 \hspace*{0.1cm}\mathrm{nm})$, $(I,\lambda)=(2.0\times 10^{14} \mathrm{W/cm}^2,800\hspace*{0.1cm} \mathrm{nm})$  and $(I,\lambda)=(3.75\times 10^{14} \mathrm{W/cm}^2,590\hspace*{0.1cm} \mathrm{nm})$, respectively.  The PADs have been plotted in a logarithmically scale. The numbers on the left top corner of each panel give the driving-field wavelength, and the letters F in the top right corner indicate that this is the full CQSFA solution. }
	\label{fig:SingleOrbPref}
\end{figure}

 For orbit 3, we also find a very good agreement between the numeric and analytic results, as shown in Fig.~\ref{fig:Orb3NoPref}. In particular, we observe that the single-orbit PADs occupy a broader momentum region for decreasing driving-field wavelength. One should note that the shape of the distributions remains similar. However, they scale with increasing frequencies. Hence, for the region of interest, longer wavelengths favor the signal along the $p_{\perp}$ axis, and reduce the region along the $p_{\parallel}$ axis for which the probability density is significant. Inclusion of the prefactor (Fig.~\ref{fig:SingleOrbPref}) locates the distributions along the $p_{\parallel}$ axis for orbit 3 and reduces the off axis probability density. However, we observe an overall decrease in the signal as the driving-field wavelength increases (see right panels in Fig.~\ref{fig:SingleOrbPref}). This agrees with the experimental findings that the spider loses relevance for higher frequencies \cite{Maharjan2006JPB}. The effect of the prefactor is less dramatic for orbits 1 and 2, and the previously discussed features remain. However, it introduces a suppression in the yield around the origin for such orbits. Examples are the widening of the PADs in the $p_{\perp}$ direction with increasing frequency and the sharp spots caused by $\mathcal{I}_{V_T}$ that exist for orbit 2 (see the left and middle columns in Fig.~\ref{fig:SingleOrbPref}).

	\subsection{The continuum propagation}
	We will now approximate the continuum propagation in order to obtain analytic expressions for Eqs.~(\ref{eq:IVc}) and (\ref{eq:Imom}). The key idea is to use approximate functions for the intermediate momenta in conjunction with the low-frequency approximation applied around a physically relevant, specific time. 
	
	In the potential integral $\mathcal{I}_{V_C}$, we will assume that the momentum in Eq.~(\ref{eq:conttraj}) is either constant or piecewise constant. This leads to the approximate expression
	\begin{equation}
	\mathcal{I}_{V_C}\approx -\sum_{j=1}^{n_j-1} \int_{t_j}^{t_{j+1}}V(\mathbf{r}_{j}(\tau ))d\tau
	\label{eq:Vcontapprox}
	\end{equation}
		\begin{equation}
		\mathbf{r}_j(\tau) \approx \int_{t_j}^{\tau}(\mathbf{p}_{j}+\mathbf{A}%
		(\tau'))d\tau'+\mathbf{c}_j,
			\label{eq:rcontapprox}
		\end{equation}
where $n_j-1$ is the number of subintervals for which the momentum $\mathbf{p}_j$ is assumed to be constant,  $t_j$ is the lower bound for these intervals and the constants $\mathbf{c}_j$  account for initial conditions that may be introduced in each subinterval. These intervals start at the real part of the ionization time, i.e.,  $t_1=t'_r$ and finish at the time $t_{n_j}=t$, $t \rightarrow \infty$. Depending on the specific orbit and on the integration interval, the times $t_j$ will carry different physical meanings, such as the time of ionization, recollision, etc.  A further approximation is to take $\mathbf{A}(\tau')=\mathbf{A}(\widetilde{t})$ in Eq.~(\ref{eq:rcontapprox}), where $\widetilde{t}$ is the orbit-specific time for which the potential integral is the most significant. In general, this is the time of closest approach between the electron and the core. However, if more than one orbit is taken into consideration, we must ensure that a common time $\widetilde{t}$ is taken so that both orbits are in the continuum.
This yields 
	\begin{equation}
	\mathbf{r}_j(\tau)\approx \mathbf{k}_{j}(\tau-t_j)+\mathbf{c}_j,
	\label{eq:rcont0}
	\end{equation}
where $\mathbf{k}_{j}=(\mathbf{p}_{j}+\mathbf{A}%
(\widetilde{t}))$.
The indefinite integral related to each term in Eq.~(\ref{eq:Vcontapprox}) reads
\begin{eqnarray}
\mathcal{I}(\Delta\tau_j)&=&-\int V(\mathbf{r}_{j}(\Delta \tau_j))\mathrm{d}\Delta\tau_j\nonumber\\
&=&-\frac{C}{|\mathbf{k}_j|}\ln
\left[
-\mathbf{k}_j \cdot (\Delta\tau_j\mathbf{k}_j+\mathbf{c}_j)
+|\mathbf{k}_j| \left| \Delta \tau_j\mathbf{k}_j +\mathbf{c}_j\right|
\right],
\label{eq:GeneralContinuum}
\end{eqnarray}
where $\Delta \tau_j=\tau-t_j$.	

A different approximation is employed to compute  the momentum correction (\ref{eq:Imom}). Thereby, we assume that, starting from a given initial time $\widetilde{t}$ whose physical meaning is orbit dependent, the difference $\pmb{\mathcal{P}}$ between a generic initial momentum $\mathbf{p}_j$ and a final momentum $\mathbf{p}_f$ is exponentially decaying. This means that the variable intermediate momentum $\mathbf{p}(\tau)$ is replaced by a fixed momentum $\mathbf{p}_j$, such that
\begin{eqnarray}
\mathcal{P}_{\parallel}(\tau)&=(p_{f\parallel}-p_{j\parallel})\exp(a_{\parallel}(\tau-\widetilde{t}))  \label{eq:expdecaypar}\\
\mathcal{P}_{\perp}(\tau)&=(p_{f\perp}-p_{j\perp})\exp(a_{\perp}(\tau-\widetilde{t}))\label{eq:expdecayperp},
\end{eqnarray}
where the coefficients $a_{\parallel}$ and  $a_{\perp}$ are computed using the assumptions specific to the problem at hand. For a monochromatic field, this yields
\begin{eqnarray}
\mathcal{I}_{\mathcal{P}}&=&
\frac{(p_{f\parallel}-p_{j\parallel})(p_{j\parallel}+3 p_{ f\parallel})}{4 a_{\parallel}}
+\frac{(p_{f\perp}-p_{j\perp})(p_{j\perp}+3 p_{f\perp})}{4 a_{\perp}}\nonumber \\
&+&\frac{2\sqrt{\up}(p_{f\parallel}-p_{j\parallel})}{a_{\parallel}^2+\omega^2}
\left(a_{\parallel} \cos(\omega \widetilde{t})-\omega \sin(\omega  \widetilde{t})\right).
\label{eq:Pintegral}
\end{eqnarray}
	\begin{figure}
		\includegraphics[width=0.8\textwidth]{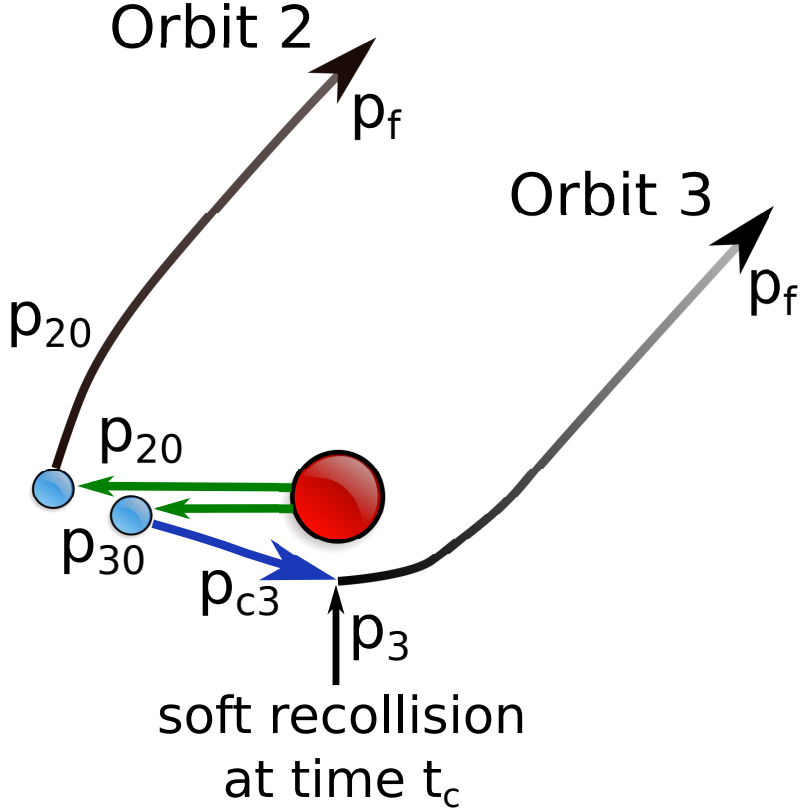}
		\caption{Schematic representation of  the approximations employed in the analytic model for orbits 2 and 3. The first part of the path in green is the tunnel exit, which is modelled as a constant momentum region as in the CQSFA. Then for orbit 2 the field dressed momentum is modelled by exponential decay to the final momentum, the fading black line. For orbit 3 the next segment in blue refers to the also constant momentum $\mathbf{p}_{3c}$ and is chosen such that a soft recollision ($z(t_c)=0$) will occur at the same time as in the CQSFA denoted $t_c$. Then orbit 3 is also described by an exponential decay from momentum $\mathbf{p}_3$ to the final momentum $\mathbf{p}_f$. Here, $\mathbf{p}_3$ is calculated by assuming an elastic collision, that scatters in electron towards the direction of the final momentum $\mathbf{p}_f$. }
		\label{fig:OrbitDiagram}
	\end{figure}
	
We will now apply the approximations discussed  above to the three main orbits that lead to intra-cycle interference. For interference to occur, they must reach the detector with the same momentum, i.e., $\mathbf{p}_{1f}=\mathbf{p}_{2f}=\mathbf{p}_{3f}=\mathbf{p}_{f}$. In all cases, we extract the tunnel exits $z_{e0}$ from Eq.~(\ref{eq:tunnelexit}) and the ionization times $t'_{er}$ from the full CQSFA according to Eqs.~(\ref{eq:t1s}), (\ref{eq:t2s}).
For orbits 1 and 2, it suffices to assume that (i) $\mathbf{p}=\mathbf{p}_{f}$ during the continuum propagation in order to calculate the potential integral $\mathcal{I}_{V_C}$; (ii) from the ionization time $t'_r$ to the end of the pulse, the momentum will tend monotonically to its final value in order to compute the momentum correction $\mathcal{I}_{\mathcal{P}}$. In contrast, for orbit 3, one must incorporate a soft collision with the core in order to reproduce the spider-like structure\footnote{We have verified that the approximations employed for orbits 1 and 2 leads to the correct behavior for orbit 3 for high photoelectron momenta  but fails to reproduce the spider-like patterns in the intermediate momentum regions. An example will be provided in Fig.~\ref{fig:spider2}}.  Specifically, we assume that the electron will follow a constant-momentum trajectory with a momentum $\mathbf{p}=\mathbf{p}_{3c}$ up to the recollision, and that it will undergo a laser driven soft collision with the core at a time $t_c$. 
 Immediately after the collision, the electron has a momentum $\mathbf{p}_3$, which is related to the collision momentum $\mathbf{p}_{3c}$ and the final momentum $\mathbf{p}_{f}$ using several approximations.  A schematic representation of the approximations used in order to compute the integrals and phase differences for orbits is plotted in Fig.~\ref{fig:OrbitDiagram}. More details are provided below.
	
	\subsubsection{Orbits 1 and 2.}
\label{subsec:orbs12}
	
Using a monochromatic driving field, and assumptions (i) and (ii), the actions $S_e$ $(e=1,2)$ associated with orbits 1 and 2 read
	\begin{eqnarray}
\hspace*{-1.5cm}	S_e(\mathbf{\tilde{p}},\mathbf{r},t,t'_e) &=&\left(\ip+ \up \right) t'_e
	+\frac{1}{2}\mathbf{p}^2_f t'_{er}
	+\frac{i}{2}\mathbf{p}^2_{e0} t'_{ei} 
	+\frac{\up}{2\omega}\sin(2\omega t'_e)  \nonumber \\
	& &+\frac{2\sqrt{\up}}{\omega}\left[
	p_{e0\parallel}\sin(\omega t'_e)-(p_{e0\parallel}-p_{f\parallel})\sin(\omega t'_{er}) 
	\right]
	-\underbrace{\int^{t'_{er}}_{t'_e}  V(\mathbf{r}_{e0}(\tau))\mathrm{d}\tau}_{\mathcal{I}^{(e)}_{V_T} } \nonumber \\
	&& \underbrace{-\frac{1}{2}\int_{t'_{er}}^{t}\pmb{\mathcal{P}}_e(\tau)\cdot (\pmb{\mathcal{P}}_e(\tau)+2\mathbf{p}_f+2\mathbf{A}(\tau))\mathrm{d}\tau}_{\mathcal{I}_{\mathcal{P}_e}}
	-\underbrace{ \int^{t}_{t'_{er}} V(\mathbf{r}_{e\mathrm{c}}(\tau))\mathrm{d}\tau}_{\mathcal{I}^{(e)}_{V_C}}.
	\label{eq:S_Analytic12} 
	\end{eqnarray}
The integrals $\mathcal{I}^{(e)}_{V_T}$, $\mathcal{I}^{(e)}_{V_C} $ and $\mathcal{I}_{\mathcal{P}_e}$ are computed as stated below. Note that making $\mathbf{p}_e$ piecewise constant eliminates the term $\mathbf{r}\cdot \dot{\mathbf{p}}$ as given by Eq.~(\ref{eq:virial}), so that there is no longer a factor 2 multiplying $\mathcal{I}^{(e)}_{V_C}$.

For $\mathcal{I}^{(e)}_{V_C}$, there will be only one interval, i.e., the lower and upper limit are $t'_{er}=\mathrm{Re}[t'_e]$ and $t$ in Eq.~(\ref{eq:Vcontapprox}).	In the approximated expression (\ref{eq:rcont0}) we take $\mathbf{k}_e=\mathbf{p}_{f}+\mathbf{A}(\widetilde{t})$, $\mathbf{c}_e=z_{e0}\hat{e}_{\parallel}$ and $t_j=t'_{er} $, where $z_{e0}$ $(e=1,2)$ are the tunnel exits for orbits 1 and 2. The time $\widetilde{t}$ is chosen as common to orbits 1 and 2. Since it must guarantee that the Coulomb effects are significant and that both orbits are in the continuum, we consider the times of closest approach for orbits 1 and 2 and take the largest of the two.

One must then compute $\mathcal{I}_e(t-t'_r)-\mathcal{I}_e(0)$, with $t \rightarrow \infty$, where $\mathcal{I }_e$ is the indefinite integral given by Eq.~(\ref{eq:GeneralContinuum}) for $e=1,2$. The lower limit reads
	\begin{equation}
	\lim\limits_{\Delta\tau_e\rightarrow0}\mathcal{I}_{e}(\Delta\tau_e)=-\frac{C}{|\mathbf{k}_e|}\ln
	\left[
	-\mathbf{k}_e\cdot\mathbf{c}_e+|\mathbf{k}_e||\mathbf{c}_e|
	\right],
	\label{eq:lower}
	\end{equation}
	while the upper limit diverges. This divergence will however cancel out for the difference $\Delta \mathcal{I}^{(12)}_{V_C}$=$\mathcal{I}^{(1)}_{V_C}$- $\mathcal{I}^{(2)}_{V_C}$, which is the quantity of interest.  The general expression for this difference is
	\begin{equation}
	\Delta\mathcal{I}^{(e e')}_{V_C} = -\frac{C}{|\mathbf{k}_{e}|}
	\ln
	\left[
	\frac{\left(-\mathbf{k}_{e'}\cdot(\mathbf{k}_{e'} \Delta\tau_{e'}+\mathbf{c}_{e'})+|\mathbf{k}_{e'}||\mathbf{k}_{e'}\Delta\tau_{e'}+\mathbf{c}_{e'}|\right)}
	{\left(-\mathbf{k}_{e}\cdot(\mathbf{k}_{e} \Delta\tau_{e}+\mathbf{c}_{e})+|\mathbf{k}_{e}||\mathbf{k}_{e}\Delta\tau_{e}+\mathbf{c}_{e}|\right)}
	\right],
	\end{equation}
	where $\Delta \tau_e=\tau-t'_{er}$. Specifically, the upper limit reads 
\begin{equation}
	\lim _{\tau \rightarrow \infty}\Delta \mathcal{I}^{(12)}_{V_C}=-\frac{2C}{|\mathbf{p}_{f}+\mathbf{A}(\widetilde{t})|}\ln\left(\frac{|z_{20}|}{|z_{10}|}\right),
	\label{eq:Couldiff12}
\end{equation}
which, together with the lower limit as stated in Eq.~(\ref{eq:lower}), is the dominant contribution to intra-cycle interference.

\begin{figure}
		\includegraphics[width=\textwidth]{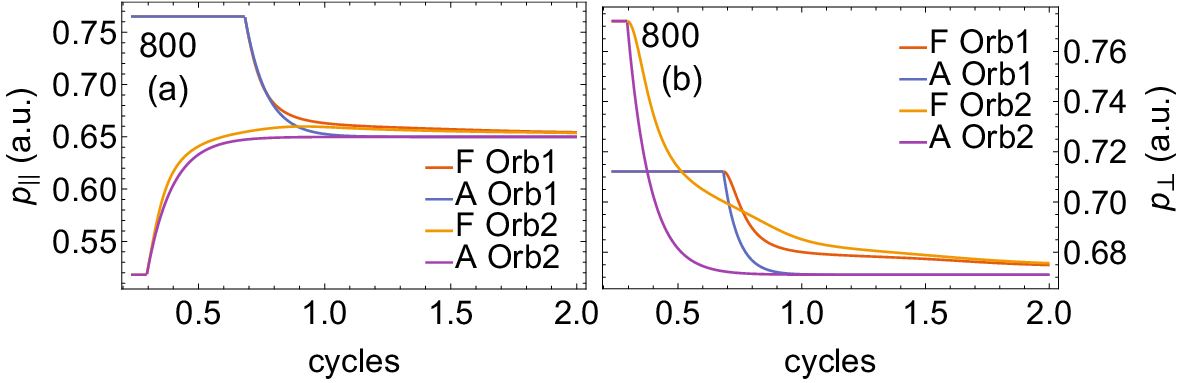}
		\caption{Exact and analytic intermediate momenta $\mathbf{p}_n$, $(n=1,2)$ for orbits one and two over two field cycles considering a field of intensity $I=2.0\times 10^{14} \hspace*{0.1cm} \mathrm{W/cm}^2$ and wavelength $\lambda=800 \hspace*{0.1cm}\mathrm{nm}$. Panels (a) and (b) give the parallel and perpendicular momentum components $p_{\parallel}$ and $p_{\perp}$, respectively. The capital letters F and A refer to the full CQSFA and the analytic approximation, respectively.}
		\label{fig:momvar}
	\end{figure}

	The momentum corrections $\mathcal{I}_{\mathcal{P}_e}$ are computed by  taking $\widetilde{t}=t'_{er}$ and $\ \mathbf{p}_e=\mathbf{p}_{e0}$ $(e=1,2)$ in Eqs.~(\ref{eq:expdecaypar}) and (\ref{eq:expdecayperp}). This is justified by the fact that, for orbits 1 and 2, the intermediate momentum tends monotonically towards the final momentum from the ionization time to the end of the pulse (see Fig.~\ref{fig:momvar}). 
	
The coefficient $a_{e \parallel}$ is evaluated at the tunnel exit and in the parallel direction using the saddle point Eq.~(\ref{eq:p-spe}) and the approximate action (\ref{eq:S_Analytic12}). This gives
	\begin{equation}
	a_{e\parallel}=-\frac{C}{z_{e0}^2 (p_{e0\parallel}-p_{f\parallel})}.
	\end{equation}
	In order to compute $a_{e\perp}$, one must bear in mind that the electron starts on the parallel axis. Thus, the right hand side of Eq.~ (\ref{eq:p-spe}) ($\nabla V(\mathbf{r}_0)\cdot \hat{e}_{\perp}$) is zero as $x(t'_r)=0$. Hence, we take the derivative with respect to time of both sides instead. This yields
\begin{equation}
a_{e\perp}=-\sqrt{\frac{C p_{e0\perp}}{|z_{e0}|^3(p_{e0\perp}-p_{f\parallel})}}.
\label{eq:aperp12}
\end{equation}
These coefficients are then used in Eq.~(\ref{eq:Pintegral}) and in the phase difference
\begin{equation}
\Delta \mathcal{I}_{\mathcal{P}_{12}}=\mathcal{I}_{\mathcal{P}_1}-\mathcal{I}_{\mathcal{P}_2}.
\label{eq:Pdiff12}
\end{equation} 
	\subsubsection{Orbit 3 and rescattering.}
	Below we discuss the approximations performed for orbit 3.  The initial momentum, used in the under-the-barrier trajectories and in $\mathcal{I}^{(3)}_{V_T}$, is $\mathbf{p}_{30}$, and the final (given) momentum is $\mathbf{p}_{f}$. 
	The continuum propagation will require two subintervals: (i) from the ionization time $t'_{3r}=\mathrm{Re}[t'_{3}]$ to a time $t_c$ for which a soft collision with the core occurs, and (ii) from the recollision time $t_c$ to the final time $t$, $t \rightarrow \infty$. 
	The time $t_c$ is calculated by solving $z(t_c)=0$ for the CQSFA, and, in agreement with the approximations in this work, is the time $\widetilde{t}$ of closest approach from the core for orbit 3. The collision being soft implies that $x(t_c)\neq 0$. We consider that, from $t'_{3r}$ to $t_c$, the perpendicular momentum component remains the same, i.e., $p_{3c\perp}=p_{30\perp}$ and the parallel component $p_{3c\parallel}$ is given by
	\begin{equation}
	\int_{t'_r}^{t_c}(p_{3c\parallel}+A(\tau))\mathrm{d}\tau +z_{30}=0.
	\label{eq:p3c}
	\end{equation}
One should note that the initial parallel momentum $\mathbf{p}_{30\parallel}$ cannot be chosen for this segment. Although the laser is the main driving force for the collision, some electron trajectories additionally require the attraction of the core to collide. Furthermore, the momentum $p_{3c\parallel}$ described by Eq.~(\ref{eq:p3c})  ensures that scattering off the core occurs at the correct time $t_c$ as determined by the CQSFA.	

Upon recollision, we assume that the electron momentum changes instantaneously from $\mathbf{p}_{3c}$ to $\mathbf{p}_3$. The latter can be fully determined using the following simplifications: (i) elastic scattering at $t_c$, i.e.,  $|\mathbf{p}_{3c}|^2 = |\mathbf{p}_{3}|^2$; (ii) the scattering angle remains the same until the end of the pulse, i.e.,  $p_{3\parallel}/p_{3\perp}=p_{f\parallel}/p_{f\perp}$. The intermediate momentum $\mathbf{p}_3$ computed as stated above will be employed in the momentum corrections $\mathcal{I}_{\mathcal{P}_3}$, but will not be used in the potential integrals $\mathcal{I}^{(3)}_{V_C}$.

The approximate expression for the action along orbit 3 reads
\begin{eqnarray}
\hspace*{-1.5cm}S_3(\mathbf{\tilde{p}},\mathbf{r},t,t'_3) &=&\left(\ip+ \up \right) t'_3
+\frac{1}{2}({p^2_{3c\parallel}} -p_{30\parallel}^2)t'_{3r}
+\frac{1}{2}(\mathbf{p}^2_{f} -{\mathbf{p}^2_{3c}})t_c
+\frac{\up}{2\omega}\sin(2\omega t'_3) \nonumber \\
&+&
\frac{2\sqrt{\up}}{\omega}\left[
p_{30\parallel}\sin(\omega t'_3)-(p_{30\parallel}-p_{3c\parallel})\sin(\omega t'_{3r}) 
-(p_{3c\parallel}-p_{f\parallel})\sin(\omega t_c)
\right] \nonumber \\
&+& \mathcal{I}_{\mathcal{P}_3}+\mathcal{I}^{(3)}_{V_T}+\mathcal{I}^{(3)}_{V_C},
\label{eq:S_Analytic3}
\end{eqnarray}
where the first two lines give a SFA-like action, and the remaining terms yield the corrections. One should note that the above-stated equation differs from Eq.~(\ref{eq:sreal}) in the sense that the change of momentum at the scattering time $t_c$ has been incorporated. This is consistent with the fact that orbit 3 lies beyond the scope of the SFA transition amplitude for direct ATI electrons \cite{Yan2010,Lai2015}. 

The tunnel integral 
\begin{equation}
\mathcal{I}^{(3)}_{V_T}=-\int^{t'_{3r}}_{t'_3}  V(\mathbf{r}_{3 0}(\tau))\mathrm{d}\tau	
\end{equation} is approximated by Eq.~(\ref{eq:ITunFinal}). The Coulomb integral $\mathcal{I}^{(3)}_{V_C}$ in the continuum must be considered within two subintervals: (i) from the ionization time $t'_r$ to the collision time $t_c$, and (ii) from the collision time $t_c$ to the final time $t$, $t \rightarrow \infty$. Explicitly,
\begin{equation}
\mathcal{I}^{(3)}_{V_C}=-\int^{t_c}_{t'_{3r}}  V(\mathbf{r}_{3c}(\tau))\mathrm{d}\tau-\int^{t}_{t_c}  V(\mathbf{r}_{3f}(\tau))\mathrm{d}\tau,
\label{eq:IVc3}
\end{equation}
where \begin{equation}
\mathbf{r}_{3c}(\tau)=\int_{t'_{er}}^{\tau}(\mathbf{p}_{3c}+\mathbf{A}(\tau'))\mathrm{d}\tau'+\underbrace{\Re[\mathbf{r}_{30}(t'_{3r})]}_{z_{30}\hat{e}_{\parallel}}
\label{eq:r3c}
\end{equation}
and
\begin{equation}
\mathbf{r}_{3\mathrm{f}}(\tau)=\int_{t_c}^{\tau}(\mathbf{p}_{f}+\mathbf{A}(\tau'))\mathrm{d}\tau'+\mathbf{r}_{3c}(t_c).
\label{eq:r3f}
\end{equation}
For both integrals we take $A(\tau') \approx A(t_c)$, as the collision time is when the contributions of the binding potential are expected to be most relevant. This gives $\mathbf{k}_{3f}=\mathbf{p}_{f}+\mathbf{A}(t_c)$ for Eq.~(\ref{eq:r3f}). In order to compute the continuum phase differences, it is convenient to rewrite Eq.~(\ref{eq:r3c}) using the assumptions stated above. Eq.~(\ref{eq:p3c}) provides us with the tunnel exit $z_{30}$, which, if combined with the parallel component of $\mathbf{r}_{3c}(\tau)$ gives
\begin{equation}
r_{3c\parallel}(\tau)=\int_{t_c}^{\tau}[p_{3c\parallel}+A(\tau ')]d\tau ' \approx [p_{3c\parallel}+A(t_c)](\tau-t_c).
\end{equation}
Constant $p_{3\perp}$ between ionization and recollision times, i.e., $p_{3c\perp}=p_{30\perp}$, then yields
\begin{equation}
\mathbf{r}_{3c}(\tau)=[p_{3c\parallel}+A(t_c)](\tau-t_c)\hat{e}_{\parallel}+p_{0\perp}(\tau-t'_r)\hat{e}_{\perp},
\end{equation}
which can be rewritten as
\begin{equation}
\mathbf{r}_{3c}(\tau)=\mathbf{k}_{3c}(\tau-t_c)+\mathbf{c}_3,
\label{eq:longwaver3c}
\end{equation}
 with $\mathbf{k}_{3c}=\mathbf{p}_{3c}+\mathbf{A}(t_c)$ and $\mathbf{c}_3=p_{30\perp}(t_c-t'_{3r})\hat{e}_{\perp}$. Similarly, 
\begin{equation}
\mathbf{r}_{3f}(\tau)=\mathbf{k}_{f}(\tau-t_c)+\mathbf{c}_3,
\label{eq:longwaver3f}
\end{equation}
with $\mathbf{k}_{f}=\mathbf{p}_{f}+\mathbf{A}(t_c)$. We will now use Eq.~(\ref{eq:GeneralContinuum}) to solve the two integrals in Eq.~(\ref{eq:IVc3}). For the first subinterval, we have 
\begin{equation}
\mathcal{I}_{\mathrm{col}}=-\int^{t_c}_{t'_{3r}}  V(\mathbf{r}_{3c}(\tau))\mathrm{d}\tau=\lim\limits_{\Delta \tau_c \rightarrow 0}\mathcal{I}_3(\Delta \tau_c)-\mathcal{I}_3(t'_{3r}-t_c).
\end{equation}
This gives
\begin{equation}
\mathcal{I}_\mathrm{col}=-\frac{C}{|\mathbf{k}_{3c}|}\ln
\left[
\frac{-\mathbf{k}_{3c}\cdot\mathbf{c}_{3c}+|\mathbf{k}_{3c}||\mathbf{c}_{3c}|}
{-\mathbf{k}_{3c}\cdot(\mathbf{k}_{3c}(t'_r-t_c)+\mathbf{c}_{3c})+|\mathbf{k}_{3c}||\mathbf{k}_{3c}(t'_r-t_c)+\mathbf{c}_{3c}|}
\right],
\end{equation}

which can be simplified further to
\begin{eqnarray}
\mathcal{I}_\mathrm{col}&=&-\frac{C}{|\mathbf{k}_{3c}|}\ln
\left[
\frac{-p_{30\perp}^2(t_c-t'_{3r})+|p_{30\perp}||k_{3c}||t_c-t'_{3r}|}
{(p_{3c\parallel}+A(t_c))^2(t_c-t'_{3r})+|k_{3c}||p_{3c\parallel}+A(t_c)||t_c-t'_{3r}|}
\right]\nonumber \\
&=&-\frac{C}{|\mathbf{k}_{3c}|}\ln
\left[
\frac{-p_{30\perp}^2+|\mathbf{k}_{3c}||p_{30\perp}|}
{(p_{3c\parallel}+A(t_c))^2+|\mathbf{k}_{3c}||p_{3c\parallel}+A(t_c)|}
\right].
\end{eqnarray}
The second integral is computed in a similar way as those in Sec.~\ref{subsec:orbs12}, with the difference that the common time will be the recollision time $t_c$ for orbit 3. Explicitly, the upper limit for 
the Coulomb phase difference between orbit 3 and one of the other two orbits, as discussed in Sec.~\ref{subsec:orbs12}, will be 
\begin{equation}
\lim\limits_{\tau \rightarrow \infty}\Delta \mathcal{I}^{(e3)}_{V_C}=-\frac{2C}{|\mathbf{k}_f|}\ln
\left(
\frac{p_{30\perp}^2 (p_{f\parallel}+A_{\parallel}(t_c))^2 (t_c-t'_{3r})^2}
{p_{f\perp}^2z_{e0}^2}
\right),
\end{equation}
with $e=1,2$. The lower limit can be computed from Eq.~(\ref{eq:lower}) directly. 
The momentum integral $\mathcal{I}_{\mathcal{P}_3}$ is computed assuming an exponential decay from the recollision time $t_c$ to the final time $t \rightarrow \infty$.  Prior to that, the momentum $\mathbf{p}_3$ is assumed to be constant and equal to $(p_{3\parallel},p_{3\perp})=(p_{3c\parallel},p_{30\perp})$ and the resulting phase shift is incorporated in the SFA-like part of the action. This means that, in Eq.~(\ref{eq:expdecaypar})-(\ref{eq:Pintegral}), $\mathbf{p}_{j}=\mathbf{p}_{3}$, which is determined according to the simplification (ii) specified above, and the closest approach time is taken as $\widetilde{t}=t_c$.  A further subtlety is that, in order to obtain the coefficient $a_{3\parallel}$ from the action (\ref{eq:S_Analytic3}), one must use the derivative of Eq.~(\ref{eq:p-spe}) as $z(t_c)=0$, so that
\begin{equation}
a_{3\parallel}=-\sqrt{\frac{C p_{3\parallel}}
	{|p_{30\perp}|^3|p_{3\parallel}-p_{3f\parallel}| (t_c-t'_{3r})}}.
\end{equation}
Finally, for the perpendicular direction,
\begin{equation}
a_{3\perp}=-\frac{C}{p_{30\perp}^2(p_{3\perp}-p_{3f\perp})(t_c-t'_{er})}.
\end{equation}
One should note that, in the above-stated equation, it was not necessary to take the time derivative of Eq.~(\ref{eq:p-spe}) as the transverse component $x(t_c) \neq 0$. Phase differences due to the momentum changes are then computed by taking 
$
\Delta \mathcal{I}_{\mathcal{P}_{e3}}=\mathcal{I}_{\mathcal{P}_e}-\mathcal{I}_{\mathcal{P}_3}$,
with $e=1,2$.

\section{Holographic structures}
\label{sec:Holographic}

\label{sec:qinterf}
\subsection{Full Comparison}
\begin{figure}
	\includegraphics[width=\textwidth]{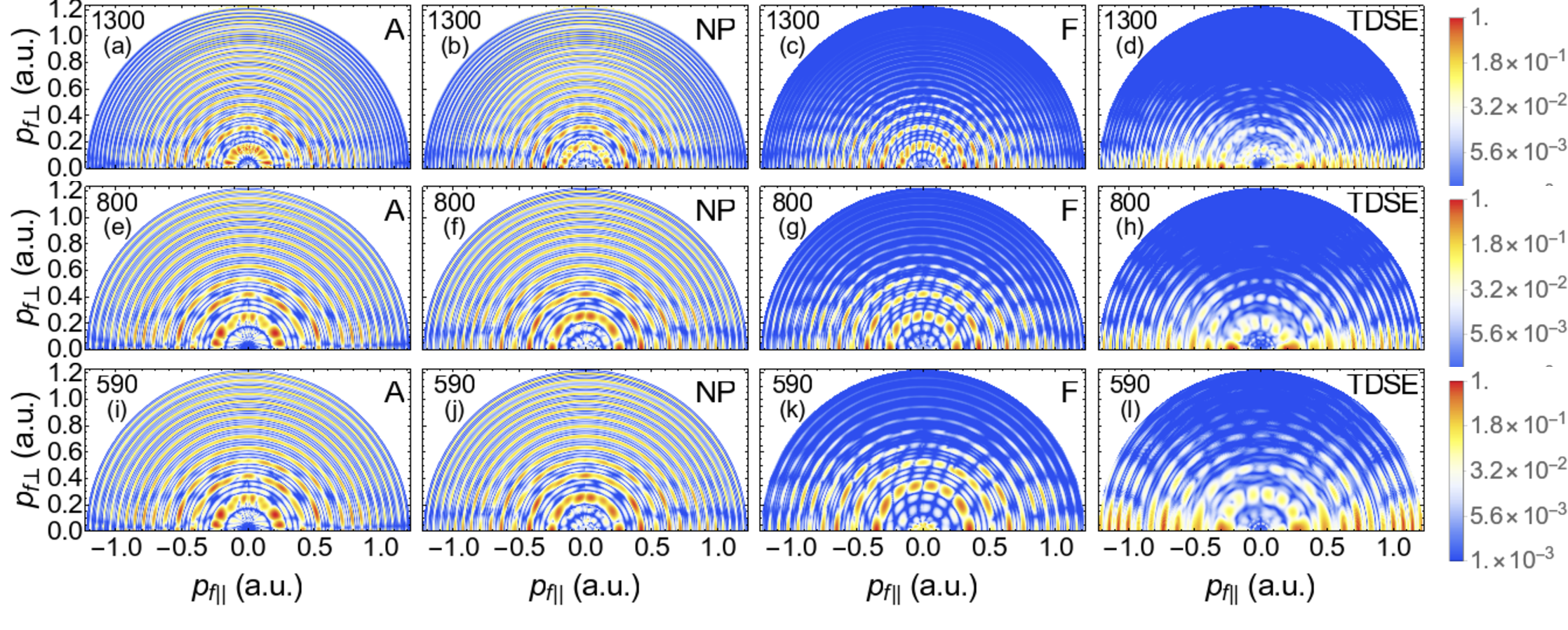}
	\caption{ATI PADs computed with the analytic condition (left column, denoted by A), the CQSFA solution without prefactors (second left column, denoted by NP), the full CQSFA solution (second right column, denoted by F) and the TDSE (right column, denoted by TDSE) for hydrogen ($I_p=0.5$ a.u.) over four driving-field cycles. A trapezium envelope was used for the TDSE results, where the flat top part was four cycles with a half-cycle ramp on and off. The lower case letters on the top left corner [panels (a) to (l)] correspond to the field parameters  $(I,\lambda)=(7.5\times 10^{13} \mathrm{W/cm}^2,1300 \hspace*{0.1cm}\mathrm{nm})$ [panels (a) to (d)], $(I,\lambda)=(2.0\times 10^{14} \mathrm{W/cm}^2,800\hspace*{0.1cm} \mathrm{nm})$ [panels (e) to (h)] and $(I,\lambda)=(3.75\times 10^{14} \mathrm{W/cm}^2,590\hspace*{0.1cm} \mathrm{nm})$ [panels (i) to (l)], where $I$ and $\lambda$ give the field intensity and wavelength, respectively. The density plots have been plotted in a logarithmic scale and normalized to the highest yield in each panel. The numbers on the top left corner of each panel give the driving-field wavelength. }
	\label{fig:Fullcomparison}
\end{figure}

In Fig.~\ref{fig:Fullcomparison}, we compare PADs computed using different means over four driving-field cycles. This includes the full CQSFA spectra with and without prefactors, the full solution of the time-dependent Schr\"odinger equation (TDSE), computed with the freely available software Qprop \cite{Qprop}, and the analytic expressions derived in the previous sections.  All PADs exhibit a myriad of patterns, including the rings caused by inter-cycle interference, the spider-like patterns near the
polarization axis that result from the interference of orbits 2 and 3, and the 
near-threshold, fan-shaped structures caused by the interference of orbits 1
and 2. 

In general, the full CQSFA  and TDSE solutions, shown in the second right [panels (c), (g) and (k)] and right [panels (d), (h) and (l)] columns, exhibit a very good agreement. However, the CQSFA underestimates the signal near the origin and the polarization axis, differs from the full TSDE solution around the $p_{\perp}$ axis and leads to different slopes for the spider.  The discrepancy near the origin may be attributed to several approximations made in the CQSFA, such as neglecting bound-state depletion and ionization pathways involving excited states. Furthermore, one assumes that the main ionization mechanism is tunnel ionization. For that reason, the Keldysh parameter $\gamma=\sqrt{I_p/(2U_p)}$ has been kept fixed and well within the tunneling regime. However, this only indicates the prevalent ionization mechanism, but it does not rule out above-the-barrier or multiphoton ionization. The agreement between the slopes of the spider-like patterns worsens for decreasing driving-field wavelength. It is quite good for 
 $\lambda=1300$ nm [panels (c), and (d)], reasonable for $\lambda=800$ nm [panels (g) and (h)] and poor for $\lambda=590$ nm. As the wavelength decreases, the TDSE slope moves away from the polarization axis, while its CQSFA counterpart remains nearly horizontal.  This is likely to be caused by the longer electron excursion amplitudes in the mid-IR regime, which increase the influence of the driving field and reduce the role of the Coulomb potential. Finally, orbit 4 has not been included in our computations and could start to play a small, but non-negligible role near the $p_{\perp}$ axis. 

In the two left panels of Fig.~\ref{fig:Fullcomparison}, we compare the full CQSFA and the analytic approximation as derived in Sec.~\ref{sec:Analytic}. This comparison can only be performed if one leaves out the prefactors, as they have not been included in the approximate expressions. They play a secondary, but important role in the PADs, by determining the relative weight between the orbits, their stability and wave-packet spreading.  This makes all PADs more uniformly distributed in momentum space, instead of concentrated around the polarization axis, and modifies the interference patterns.  In the absence of prefactors, the CSQFA fringes appear more blurred and blotched, and there is good agreement with the analytic expressions for a wide range of driving-field parameters.  Thus, the additional approximations carried out in the previous section can be used in analysing specific holographic patterns more closely.

\subsection{Coulomb effects in intra-cycle interference}
\label{subsec:fan}

\begin{figure}
	\includegraphics[width=\textwidth]{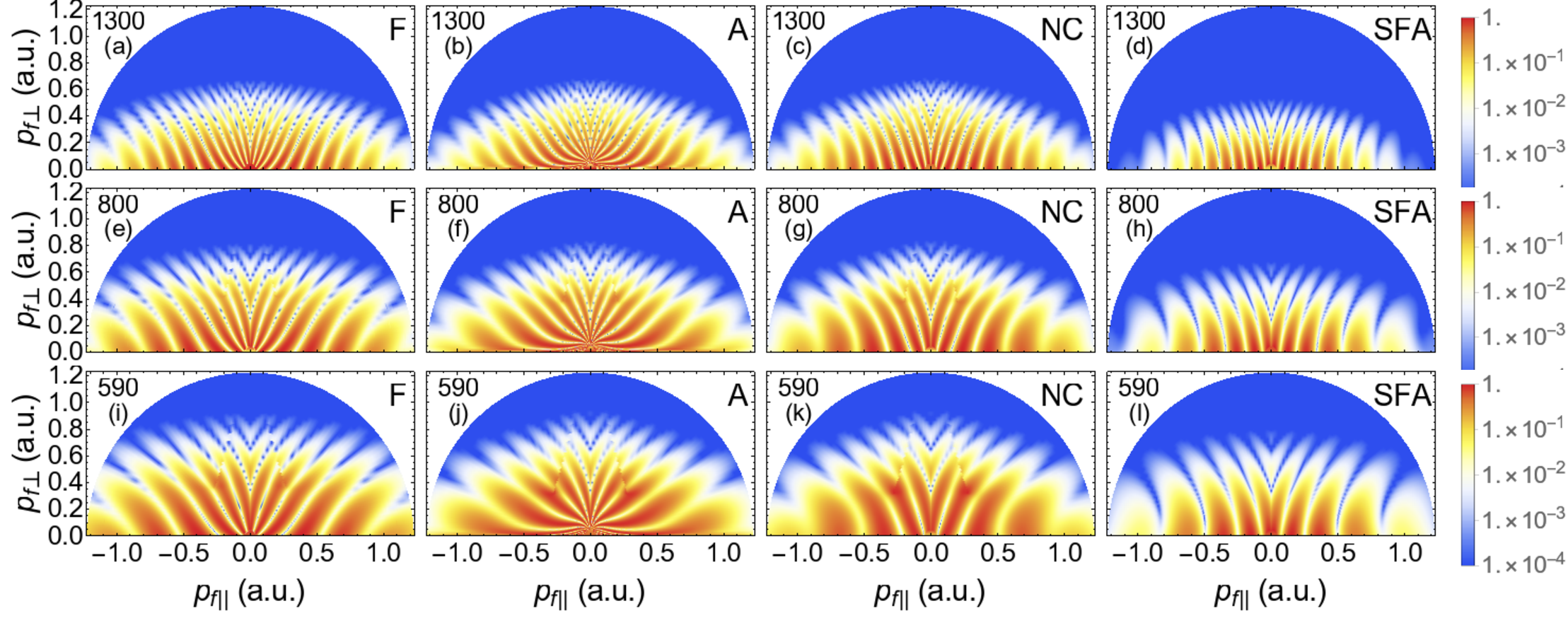}
	\caption{Fan-shaped holographic structures computed for Hydrogen using orbits 1 and 2 and symmetrizing upon $p_{\parallel}=0$ so that $\Re[t_1-t_2]$ is smaller than or at most equal to half a field cycle. The left columns [panels (a), (e) and (i)] provide the numerical CQSFA solution without prefactor, the second left column [panels (b), (f) and (j)] show the analytical approximations, the second right column [panels (c), (g) and (k)] display the analytical  model without the Coulomb phases $\mathcal{I}_{V_C}$, and the far right column [panels (d), (h) and (l)] show the equivalent patterns for the SFA-like term in the action. The field parameters for the first, second and third row are the same as in Fig.~\ref{fig:Fullcomparison}. The probability densities have been normalized to the maximum yield in each panel and plotted in a logarithmic scale. The upper-case letters F, A, NC and SFA on the top right corners of each panel mean full CQSFA, analytic CQSFA, CQSFA with no Coulomb integral and the SFA-like part of the transition amplitude, where the integral corrections are not included, respectively. The numbers on the top left corner of each panel indicate the driving-field wavelength. }
	\label{fig:fan1}
\end{figure}
We will next employ the analytic approximations to assess what influence the propagation integrals $\mathcal{I}_{V_C}$ and $\mathcal{I}_{\mathcal{P}}$, in addition to the SFA-like terms, have on intra-cycle interference patterns. 
Fig.~\ref{fig:fan1} displays the fan-shaped structure, which, in previous work, was shown to result from the intra-cycle interference of types 1 and 2 orbits \cite{Lai2017,Maxwell2017}, provided that $\mathrm{Re}[t_{2c}-t_{1c}]\leq \pi/\omega$. The figure shows that the analytic model overestimates the diverging behavior of this structure due to the Coulomb phase, in comparison to the full CQSFA for a wide range of field parameters. This is expected, as it has been constructed around the times $\widetilde{t}$ for which the Coulomb potential is most important, whose long tail causes the fringes to diverge.  A legitimate question is where this influence is the most critical: is it via the Coulomb phase difference (\ref{eq:Couldiff12}) or via the momentum corrections (\ref{eq:Pdiff12})? In Figs.~\ref{fig:fan1}(c), (g) and (k), we remove the Coulomb phase difference from the analytic expressions, and find that the slope of the distributions changes considerably. Furthermore, the Coulomb phase causes a narrowing of the fringes near the origin, where the affect of the Coulomb potential is the largest, which is lost when this term is removed. This can be observed to lesser extent in the full CQSFA plots, Figs.~\ref{fig:fan1}(a), (e) and (i). Still, both the momentum and Coulomb integrals contribute as the PAD computed using Eq. (\ref{eq:S_Analytic12}) without such integrals, displayed in the far right panels of the figure, are markedly different. This shows that all corrections are important in forming the fan, but that $\mathcal{I}^{(12)}_{V_C}$ is the most important contribution. 

\begin{figure}
	\includegraphics[width=\textwidth]{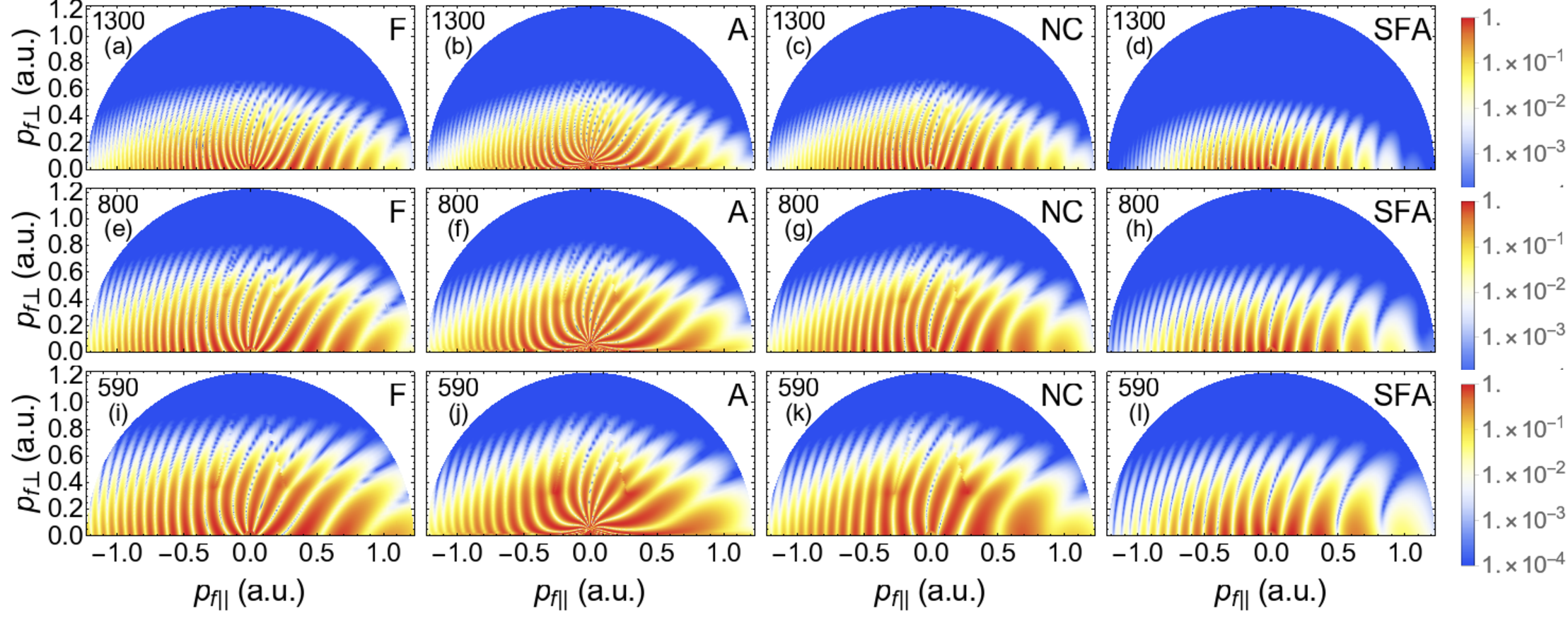}
	\caption{Holographic structures stemming from the interference of orbits 1 and 2 computed without symmetrization and relaxing the restriction upon the ionization times $t_1$ and $t_2$. We use the same field and atomic parameters, and the same notation as in Fig.~\ref{fig:fan1}. The probability densities have been normalized to the maximum yield in each panel and plotted in a logarithmic scale. }
	\label{fig:fan2}
\end{figure}
This situation persists if the restriction $\mathrm{Re}[t_{2c}-t_{1c}]\leq \pi/\omega$ is relaxed and other types of intra-cycle interference between orbit 1 and 2 are present. This can be seen in Fig.~\ref{fig:fan2}, which shows that the absence of the Coulomb phase $\mathcal{I}^{(12)}_{V_C}$ causes the interference patterns to become much closer to those obtained with the SFA.  Overall, we also see that the fringes become thicker as the driving-field wavelength decreases. Physically, this is consistent with the fact that, for longer wavelengths, the electron excursion lengths in the continuum are larger. This clearly plays a role in increasing the phase difference as orbits 1 and 2 start in different half cycles of the field. 

\begin{figure}
	\includegraphics[width=\textwidth]{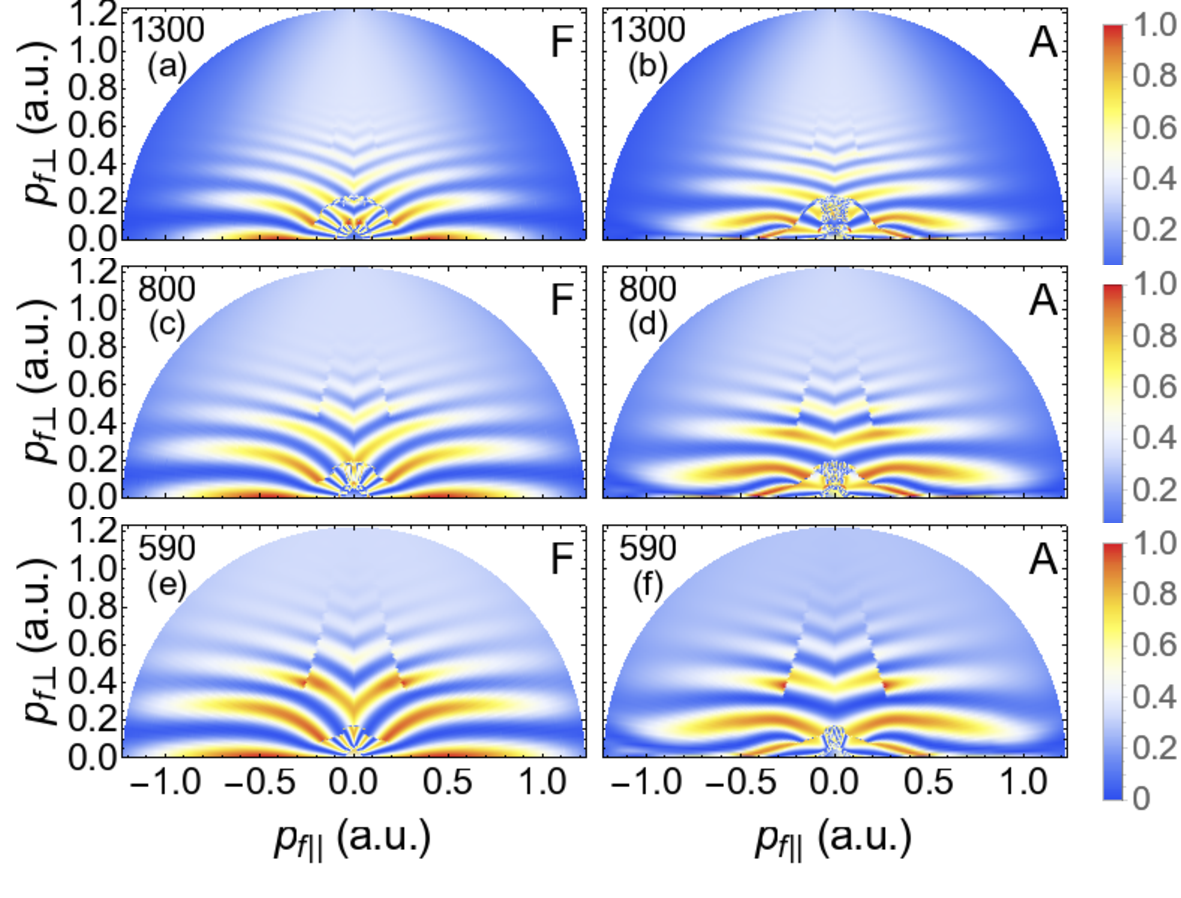}
	\caption{Spider-like structures stemming from the interference computed for the same field and atomic parameters, no prefactors, and using the same notation as in Figs.~\ref{fig:fan1} and \ref{fig:fan2}. The left and the right column have been computed using the full CQSFA and its analytical counterpart, respectively. This is indicated by the capital letters F and A in the top right corner of each panel. The probability densities have been normalized to the maximum yield in each panel and plotted in a linear scale. }
	\label{fig:spider1}
\end{figure}

Fig.~\ref{fig:spider1} shows the spider-like structures computed with the full and analytic CQSFA. Overall, we see that the slope of the full solution is nearly horizontal, while the slope of the analytic solution bends slightly upwards.  This is consistent with the fact that the upward bending is caused by the Coulomb phases in the continuum, which are overestimated in the analytic model. Interestingly, the fringe spacing changes little with the driving-field wavelength. This is due to the fact that both orbits start in the same half cycle of the field. Thus, the phase difference between them does not change much even if the electron has longer excursion amplitudes.

 Another noteworthy feature is that, near the origin, there are secondary spider-like structures in the full CQSFA, which are associated with multiple scattering events. They are particularly clear in Fig.~\ref{fig:spider1}(a) for a wavelength of 1300 nm. This can be associated with number field cycles before rescattering, 1, 3 and 5 cycles relate to the outer, inner and ``inner-inner'' spider patterns, respectively \cite{Hickstein2012,Pisanty2016,Xie2016}. The splitting is partially recovered in the analytic model, but cannot be fully accounted for as it only allows one `soft-scattering'/close return of the electron. Such soft scattering trajectories have been directly related to the low energy structure (LES) \cite{Kaestner2012JPB,Pisanty2016,Becker2015JPB}. 
 
\begin{figure}
	\includegraphics[width=\textwidth]{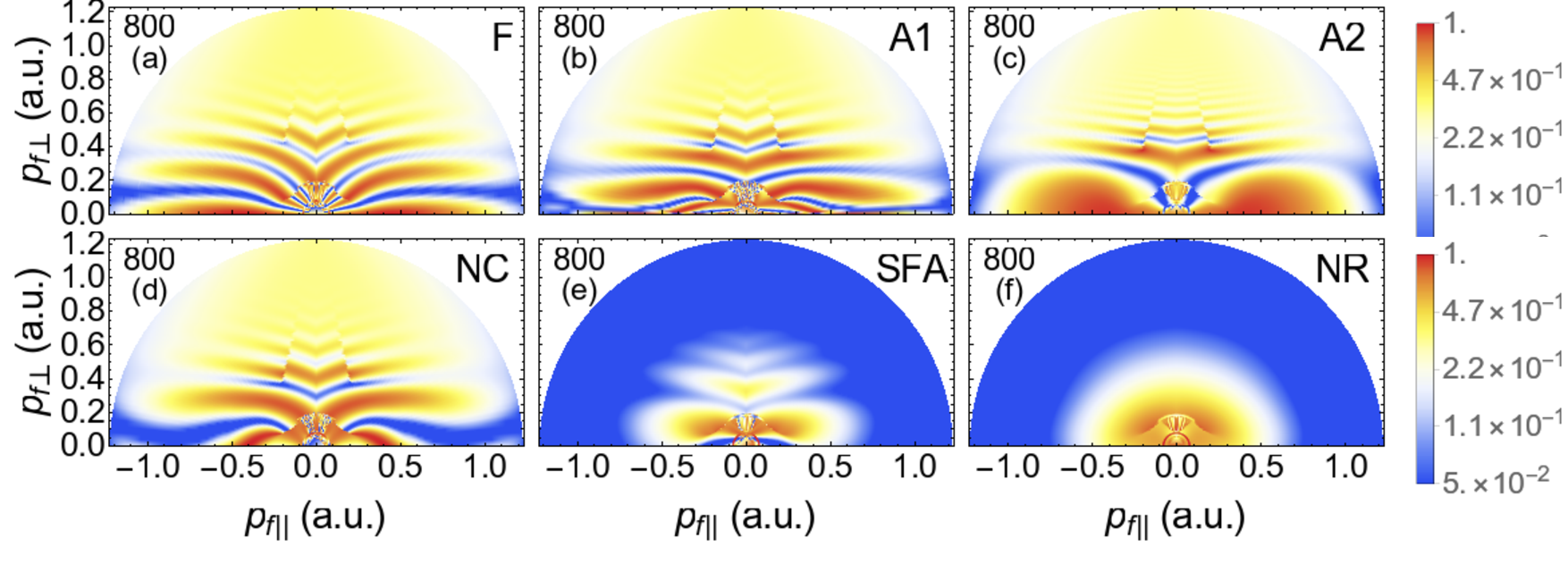}
	\caption{Spider-like structures computed for Hydrogen in a field of intensity $I=2.0\times 10^{14} \mathrm{W/cm}^2$ and wavelength $\lambda=800\hspace*{0.1cm} \mathrm{nm}$ using the full CQSFA [panel (a); indicated by F], the analytic CQSFA model with and without rescattering [panels (b) and (c); indicated by A1 and A2, respectively], the CQSFA without the Coulomb phases $\mathcal{I}_{V_C}$[panel (d); indicated by NC], the SFA-like part of the action, for which the potential and momentum integrals in the have been neglected in the continuum propagation [panel (e); indicated by SFA] and the SFA-like part of the action without rescattering [panel (f); indicated by NR]. The probability densities have been normalized to the maximum yield in each panel and plotted in a logarithmic scale.  }
	\label{fig:spider2}
\end{figure}

Fig.~\ref{fig:spider2} provides additional insight on how the Coulomb potential affects the spider. Its influence occurs in three main ways: (i)  It contributes to the Coulomb phases and to the phase difference $\mathcal{I}^{(23)}_{V_C}$; (ii) it accelerates the electron, which within our model is taken into account in the momentum integral $\mathcal{I}_{\mathcal{P}}$ and the phase difference $\mathcal{I}_{\mathcal{P}_{23}}$; (iii) it causes the electron along orbit 3 to rescatter with the core. The main effect of the Coulomb phase in (i) is to bring the fringes of the spider upwards. This can be readily seen by comparing Figs.~\ref{fig:spider2}(b) and (d), for which these integrals are present and absent, respectively. If the influence of the Coulomb potential is accounted for only as (ii) and (iii) the spider fringes bend downwards, and even cross the $p_{\parallel}$ axis. Fig.~\ref{fig:spider2}(c) models orbit 3 within the CQSFA in a similar way as for orbits 1 and 2, i.e., incorporating the Coulomb potential but without rescattering. In this case, the fringes near the axis and the central part of the spider vanish. This shows that considering the binding potential only via (i) and (ii) does not suffice for a correct description of orbit 3. Furthermore, the spider will only extend towards high photoelectron momenta if the acceleration by the potential as described in (ii) is incorporated. This is clearly seen in Fig.~\ref{fig:spider2}(e), for which both the Coulomb and the momentum integral are absent. In this case, only the central part of the spider is present, and the photoelectron energy extends to roughly $2U_p=0.88$ a.u., which is the direct ATI cutoff as given by the SFA. Finally, if neither rescattering for orbit 3 nor  the corrections $\mathcal{I}_{V_C}$ and $\mathcal{I}_{\mathcal{P}}$ are present, the distribution resembles what is obtained for a single-orbit direct ATI PAD in the standard SFA, i.e., a single peak around $(p_{\parallel},p_{\perp})=(0,0)$, which extends to the maximum energy of around $2U_p$. Physically, this could be understood as an SFA-like model with two long orbits, which carry slightly different ionization times and momenta. This very small difference would lead to very thick interference fringes, which may lie beyond the cutoff energy.

\section{Conclusions}

\label{sec:conclusions}
In this work, we provide analytic expressions for Coulomb corrected above-threshold ionization (ATI) dynamics based on the previously developed Coulomb quantum orbit strong-field approximation (CQSFA) \cite{Lai2015,Lai2017,Maxwell2017},  which allow a direct computation of quantum-interference patterns in photoelectron angular distributions (PADs). This approach is more refined than the analytic methods existing in the literature, as it includes the Coulomb potential in the ionization and continuum propagation dynamics. The former is important in determining the shapes of the electron-momentum distributions, and the latter allow us to reproduce patterns commonly encountered in photoelectron holography, such as the fan- and spider-like structures. In the ionization dynamics, the main piece of information are the times and the momenta with which the electron reaches the continuum, and how this alters the electron momentum distribution. In the continuum dynamics, the influence of the Coulomb potential manifests itself as (a) a Coulomb phase, which is accumulated during the electron propagation; (b) phase differences due to changes in the electron momentum caused by the residual potential, from the instance of ionization to the time at which it reaches the detector; (c) in some cases, rescattering does play a role and must be incorporated. This goes beyond most analytic models for holographic photoelectron structures, which are fully classical \cite{Hickstein2012,Huismanset2010Science,Xie2016} and/or are SFA-based include at most hard collisions \cite{Becker2015JPB,LiOpticsExpress2016}. More sophisticated models focus on the low-energy structures, but do not aim at reproducing holographic patterns \cite{Pisanty2016,Kaestner2012JPB}.

In contrast,  we incorporate the effects (a)--(c)  in the semiclassical action, which is directly used to computed the PADs and photoelectron patterns. 
Key approximations consist in assuming that the intermediate electron momenta are piecewise constant  when computing the Coulomb phase (a), and monotonically decaying towards its final value when computing the momentum corrections (b). We also expand the external field around the times for which the electron is closest to the core, which are determined from the numerical solution of the CQSFA.

Overall, we reproduce key features observed in intra- and intercycle interference, and obtain a good agreement with the CQSFA, provided the prefactors are neglected. The latter include further momentum bias due to the shapes of the initial bound-states, wave-packet spreading and modify the stability of each type of orbit. We employ a direct orbit, and two types of forward deflected electron orbits, which have been first identified in \cite{Yan2010} within Coulomb corrected strong-field models. These orbits have also been used in our previous work to construct the CQSFA transition amplitudes \cite{Lai2015,Lai2017,Maxwell2017}. However, in contrast to \cite{Yan2010}, in which $10^8-10^9$ contributed trajectories are needed to reproduce quantum interference features, in our model it suffices to consider one trajectory of each type. These three trajectories are also used in the analytic model. Orbits 1 and 2 exist in the standard strong-field approximation (SFA), for which the influence of the Coulomb potential is neglected in the continuum, while orbit 3 requires the residual Coulomb potential to be present \cite{Yan2010,Lai2015}. It behaves in a similar way to the forward scattered orbit present in the SFA model of high-order ATI \cite{Becker2015JPB, LiOpticsExpress2016}.

Apart from considerably decreasing the numerical effort, the analytic model allows a closer look at how the holographic structures form. For instance, in previous work, we have shown that the fan-shaped structure that forms near the ionization threshold stems from the interference of types 1 and 2 orbits. The fan arises due to an angle- and momentum dependent distortion caused by the Coulomb potential, which is maximal close to the polarization axis \cite{Lai2017,Maxwell2017}. An open question was, however, whether this distortion occurred due to the Coulomb phase or the momentum changes caused by the Coulomb potential. In the present work, we find that all Coulomb corrections contribute to the fan. However, the most dramatic effect is caused by the Coulomb phase given by the integrals $\mathcal{I}_{V_C}$, which acts to both straighten and narrow (near the origin) the fringes to give the characteristic fan shape. This is also the case for other types of intra-cycle interference involving orbits 1 and 2, which are less prominent and thus mostly overlooked in experiments. Our analytic computations also show that, when modelling orbits 1 and 2, it suffices to include corrections around a model which, in the limit of vanishing Coulomb potential, tends to the SFA without rescattering. This is consistent with the fact that orbits 1 and 2 have well-known SFA counterparts \cite{Yan2010,Lai2015} and tend to the SFA in the limit of very large photoelectron momenta \cite{Maxwell2017}.

Another widely studied holographic structure is the spider-like pattern that forms near the polarization axis and extends to very high momenta. This structure is caused by the interference of orbits 2 and 3. The present results show that the spider requires an appreciable acceleration of the electron in the continuum, the Coulomb phase and, above all, rescattering for orbit 3. In fact, analytical Coulomb corrected models similar to those developed by us for orbits 1 and 2 fail to reproduce this structure (see Fig.\ref{fig:spider2}). It was necessary to assume an abrupt momentum change at a rescattering time $t_c$, which led to a very distinct transition amplitude (Eq.~(\ref{eq:S_Analytic3})). In the limit of vanishing binding potential, Eq.~(\ref{eq:S_Analytic3}) does not tend to the direct SFA. This is supported by the fact that an electron along orbit 3 gets much closer to the core and is accelerated for a longer time than for the remaining orbits. Furthermore, orbit 3 does not have an SFA counterpart in direct ATI nor exhibits any high-energy limit that can be traced back to the direct SFA \cite{Maxwell2017}. However, there is some evidence that it could be approximated by a forward scattered SFA orbit in high-order ATI \cite{Becker2015JPB}. It is indeed noteworthy that, in the full CQSFA, the distinction between direct and rescattered electrons is blurred. In contrast, the assumptions made upon the intermediate momenta in order to compute the corrections used in this work provide a higher degree of control over the presence, absence or nature of the rescattering events taking place. Hence, we can extract the importance of soft rescattering in orbit 3, despite the fact that this orbit can exhibit behaviour that varies between deflection and hard scattering.

Interestingly, the full CQSFA takes into account multiple scattering, which is left out in the analytic effect. This causes the spider-like fringes to split in the low momentum region, leading to several inner spiders. This structures have been reported in \cite{Hickstein2012}, and are more visible for longer wavelengths. Our results also indicate that the Coulomb phase in the continuum is underestimated in the full CQSFA, especially for shorter driving-field wavelengths. This can be seen in the slope of the spider-like structure, which is strongly influenced by the Coulomb phase. For the CQSFA, the fringes forming the spider are nearly horizontal for all the parameters used, while in their TDSE counterparts the slopes in the mid-IR regime are in agreement with the CQSFA, but increase with the driving-field frequency. Physically, this is related to the fact that, the higher the frequency is, the smaller the electron excursion amplitude in the continuum will be. This means that the electron will spend more time near the core. An increase in the slope is also observed for the analytical CQSFA model, which clearly overestimates the Coulomb phase by expanding around the times of closest approach to the core [see Figs.~\ref{fig:spider1} and \ref{fig:spider2}]. 

 A shortcoming of our approach is that, in its current form, it is not a stand-alone model, as it uses the closest-approach times and initial momenta determined from the CQSFA. It is hence desirable to find an alternative, consistent criterium for determining such times and momenta. Another shortcoming is that the regularisation procedure allows from some freedom in the normalisation of the orbits due to the improper limit. A preferable regularisation procedure would not have this freedom. Further important issues are how to incorporate multiple rescattering, and to establish a direct connection between the CQSFA, the analytic model and the rescattered ATI transition amplitude computed using the SFA. Nonetheless, the analytic approach discussed in this work provides deeper insight into how holographic structures form, and yields a consistent and numerically much cheaper way of computing ATI PADs in the presence of the residual binding potential. This may be useful for computing Coulomb corrected probability distributions for more complex systems, with many degrees of freedom and more than one electron. 

We would like to thank the UK EPSRC (grant EP/J019240/1) for financial support and Richard Juggins for useful discussions.

\end{document}